\documentclass[11pt,preprint,longabstract]{aastex}
\usepackage{amsmath}

\defcitealias{pri08}{PHS}
\defcitealias{sul06b}{S06}

\shortauthors{Perrett et al.}
\shorttitle{Volumteric SN Ia Rates from the SNLS}

\def\runit{\ensuremath{\times10^{-4}\,\mathrm{SNe\,yr}^{-1}\,\mathrm{Mpc}^{-3}}}
\def\Aunit{\ensuremath{\times10^{-14}\,\mathrm{SNe\,yr}^{-1}\,M_\odot^{-1}}}
\def\Bunit{\ensuremath{\times10^{-4}\,\mathrm{SNe\,yr}^{-1}\,(M_\odot\,\mathrm{yr}^{-1})^{-1}}}
\def\mstellar{\ensuremath{M_{\mathrm{stellar}}}}
\def\mstar{\ensuremath{M_{\ast}}}
\def\msun{\ensuremath{\mathrm{M}_{\mathrm{\odot}}}}
\def\snr{\ensuremath{\mathrm{SNR}_{\mathrm{Ia}}}}

\begin{document}

\title{Evolution in the Volumetric Type Ia Supernova Rate from the Supernova
  Legacy Survey\footnotemark[1]}

\author{
K.~Perrett\altaffilmark{1,2},
M.~Sullivan\altaffilmark{3},
A.~Conley\altaffilmark{4},
S.~Gonz\'alez-Gait\'an\altaffilmark{1},
R.~Carlberg\altaffilmark{1},
D.~Fouchez\altaffilmark{5},
P.~Ripoche\altaffilmark{19,6},
J.~D.~Neill\altaffilmark{7},
P.~Astier\altaffilmark{6},
D.~Balam\altaffilmark{8},
C.~Balland\altaffilmark{6,9},
S.~Basa\altaffilmark{10},
J.~Guy\altaffilmark{6},
D.~Hardin\altaffilmark{6},
I.~M.~Hook\altaffilmark{3,11}
D.~A.~Howell\altaffilmark{12,13}
R.~Pain\altaffilmark{6}, 
N. Palanque-Delabrouille\altaffilmark{14},
C.~Pritchet\altaffilmark{15}, 
N.~Regnault\altaffilmark{6},
J.~Rich\altaffilmark{14},
V. Ruhlmann-Kleider\altaffilmark{14},
S.~Baumont\altaffilmark{6,16},
C.~Lidman\altaffilmark{17},
S.~Perlmutter\altaffilmark{18,19},
E.S.~Walker\altaffilmark{20}
}

\altaffiltext{1}{Department of Astronomy and Astrophysics, University
  of Toronto, 50 St. George Street, Toronto, ON, M5S 3H4, Canada}

\altaffiltext{2}{Network Information Operations, DRDC Ottawa, 3701
  Carling Avenue, Ottawa, ON, K1A~0Z4, Canada}

\altaffiltext{3}{Department of Physics (Astrophysics), University of
  Oxford, DWB, Keble Road, Oxford OX1 3RH, UK}

\altaffiltext{4}{Center for Astrophysics and Space Astronomy, University of
  Colorado, 593 UCB, Boulder, CO, 80309-0593, USA}

\altaffiltext{5}{CPPM, CNRS-IN2P3 and University Aix Marseille II,
  Case 907, 13288 Marseille cedex 9, France}

\altaffiltext{6}{LPNHE, Universit\'e Pierre et Marie Curie Paris 6,
  Universit\'e Paris Diderot Paris 7, CNRS-IN2P3, 4 place Jussieu,
  75005 Paris, France}

\altaffiltext{7}{California Institute of Technology, 1200 East
  California Blvd., Pasadena, CA, 91125, USA}

\altaffiltext{8}{Dominion Astrophysical Observatory, Herzberg
  Institute of Astrophysics, 5071 West Saanich Road, Victoria, BC,
  V9E~2E7, Canada}

\altaffiltext{9}{Universit\'e Paris 11, Orsay, F-91405, France}

\altaffiltext{10}{Laboratoire d'Astrophysique de Marseille, P\^{o}le de
  l'Etoile Site de Ch\^{a}teau-Gombert, 38, rue Fr\'{e}d\'{e}ric
  Joliot-Curie, 13388 Marseille cedex 13, France}

\altaffiltext{11}{INAF, Osservatorio Astronomico di Roma, via Frascati
  33, 00040 Monteporzio (RM), Italy}

\altaffiltext{12}{Las Cumbres Observatory Global Telescope Network,
  6740 Cortona Dr., Suite 102, Goleta, CA 93117, USA}

\altaffiltext{13}{Department of Physics, University of California,
  Santa Barbara, Broida Hall, Mail Code 9530, Santa Barbara, CA
  93106-9530, USA}

\altaffiltext{14}{DSM/IRFU/SPP, CEA-Saclay, F-91191 Gif-sur-Yvette,
  France}

\altaffiltext{15}{Department of Physics \& Astronomy, University of
  Victoria, PO Box 3055, Stn CSC, Victoria, BC, V8W~3P6, Canada}

\altaffiltext{16}{LPSC, CNRS-IN2P3, 53 rue des Martyrs, 38026 Grenoble
  Cedex, France}

\altaffiltext{17}{Australian Astronomical Observatory, P.O. Box 296, Epping,
  NSW 1710, Australia}

\altaffiltext{18}{Department of Physics, University of California, 366
  LeConte Hall MC 7300, Berkeley, CA 94720-7300, USA}

\altaffiltext{19}{Lawrence Berkeley National Laboratory, Mail Stop
  50-232, Lawrence Berkeley National Laboratory, 1 Cyclotron Road,
  Berkeley, CA 94720, USA}

\altaffiltext{20}{Scuola Normale Superiore, Piazza dei Cavalieri 7,
  56126 Pisa, Italy}

\email{perrett@astro.utoronto.ca,sullivan@astro.ox.ac.uk}

\begin{abstract}
  We present a measurement of the volumetric Type Ia supernova (SN~Ia)
  rate (\snr) as a function of redshift for the first four years of
  data from the Canada-France-Hawaii Telescope (CFHT) Supernova Legacy
  Survey (SNLS).  This analysis includes $286$
  spectroscopically confirmed and more than $400$ additional
  photometrically identified SNe Ia within the redshift range $0.1\leq
  z\leq 1.1$.  The volumetric \snr\ evolution is consistent with a
  rise to $z\sim 1.0$ that follows a power-law of the form
  (1+$z$)$^{\alpha}$, with $\alpha={2.11\pm 0.28}$.  This evolutionary
  trend in the SNLS rates is slightly shallower than that of the
  cosmic star-formation history over the same redshift range. We
  combine the SNLS rate measurements with those from other surveys
  that complement the SNLS redshift range, and fit various simple SN
  Ia delay-time distribution (DTD) models to the combined data. A
  simple power-law model for the DTD (i.e., $\propto t^{-\beta}$)
  yields values from $\beta=0.98\pm0.05$ to $\beta=1.15\pm0.08$
  depending on the parameterization of the cosmic star formation
  history.  A two-component model, where \snr\ is dependent on stellar
  mass (\mstellar) and star formation rate (SFR) as $\snr(z)=A\times
  \mstellar(z) + B\times\mathrm{SFR}(z)$, yields the coefficients
  $A=(1.9\pm 0.1)\Aunit$ and $B=(3.3\pm 0.2)\Bunit$. More general
  two-component models also fit the data well, but single Gaussian or
  exponential DTDs provide significantly poorer matches.  Finally, we
  split the SNLS sample into two populations by the light curve width
  (stretch), and show that the general behavior in the rates of
  faster-declining SNe~Ia ($0.8\leq s < 1.0$) is similar, within our
  measurement errors, to that of the slower objects ($1.0\leq s <
  1.3$) out to $z\sim 0.8$.
\end{abstract}


\section{Introduction}

Type Ia supernova (SN~Ia) explosions play a critical role in
regulating chemical evolution through the cycling of matter in
galaxies.  As supernovae (SNe) are the primary contributors of heavy
elements in the universe, observed variations in their rates with
redshift provide a diagnostic of metal enrichment over a cosmological
timeline.  The frequency of these events and the processes involved
provide important constraints on theories of stellar evolution.

SNe~Ia are thought to originate from the thermonuclear explosion of
carbon-oxygen white dwarfs that approach the Chandrasekhar mass via
accretion of material from a binary companion \citep[for reviews,
see][]{hn00,how11}.  This process can result in a significant ``delay
time'' between star formation and SN explosion, depending on the
nature of the progenitor system \citep{mad98,gre05}.  The SN~Ia
volumetric rate (\snr) evolution therefore represents a convolution of
the cosmic star-formation history with a delay-time distribution
(DTD).  As such, measuring the global rate of SN~Ia events as a
function of redshift may be useful for constraining possible DTDs and,
ultimately, progenitor models -- the detailed physics of SNe Ia
remains poorly understood, with several possible evolutionary paths
\citep[e.g.][]{bra95,Liv00}.

One complication for rates studies is that many SN surveys at low
redshifts are galaxy-targeted, counting discoveries in a select sample
of galaxies and converting to a volumetric rate by assuming a galaxy
luminosity function.  This method can be susceptible to systematic
errors if it preferentially samples the bright end of the galaxy
luminosity function, biasing toward SNe in more massive, or brighter,
galaxies \citep[see, e.g.,][]{sul10}.  Since many SN Ia properties are
correlated with their hosts, the recovered rates may then not be
representative of all types of SNe~Ia.  A second type of SN survey
involves making repeat observations of pre-defined fields in a
``rolling search'', to find and follow SNe in specific volumes of sky
over a period of time.  Such surveys minimize the influence of host
bias, but still suffer from Malmquist bias and other selection
effects.  It is reasonably straight forward --- although often
computationally expensive --- to compensate for the observational
biases within rolling searches.

The advent of these wide-field rolling surveys has significantly
enhanced SN statistics at cosmological distances.  The Supernova
Legacy Survey (SNLS) in particular has contributed a large sample of
Type Ia SNe out to redshifts of $z\sim1.05$ \citep{guy10}.  Although
its primary goal is to assemble a sample of SNe~Ia to constrain the
cosmological parameters \citep[e.g.][]{ast06,sul11}, the SNLS is also
ideal for studies of SN rates \citep[][]{nei06,baz09}. The SNLS is a
rolling high-redshift search, with repeat multi-color imaging in four
target fields over five years and as such has consistent and
well-defined survey characteristics, along with significant follow-up
spectroscopy. However, due to the selection effects (including
incomplete spectroscopic follow-up) and other systematic errors, such
as contamination and photometric redshift errors, present in any SN
survey, a detailed understanding of internal biases is necessary for
accurate rate calculations.

\begin{figure}
\plotone{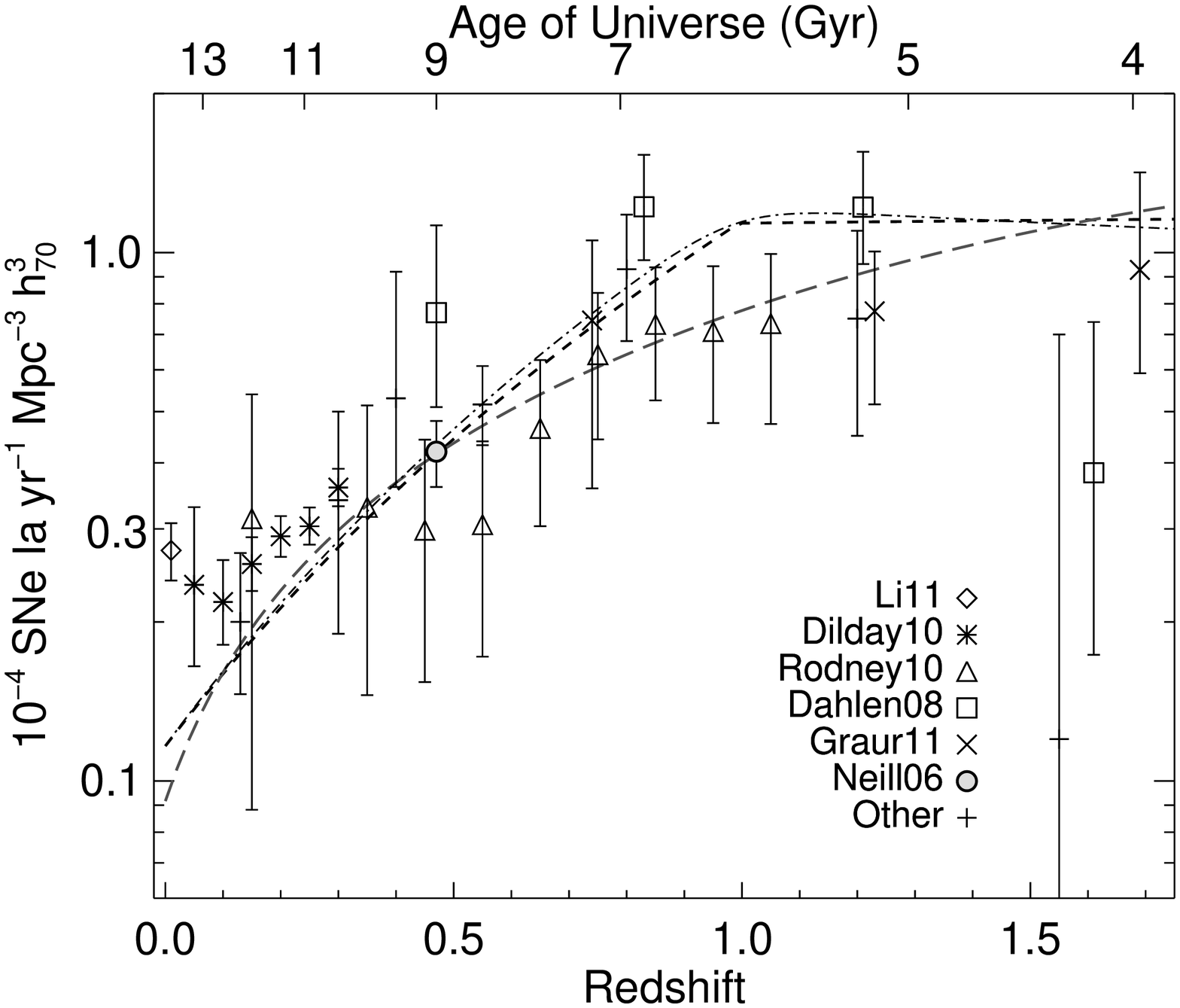}
\caption{Volumetric SN~Ia rates as a function of redshift from various
  previous studies, taken from
  \citet{li11b,dil10,rod10,dah08,gra11,nei06}.  Additional individual
  rates ($+$) include, in order of increasing redshift:
  \citet{bla04,bot08,kuz08}. Values are plotted as published, with the
  exception of a correction to the cosmology used in this paper. As a
  comparison, the lines shows the evolution of various model cosmic
  star-formation histories from \citet[][piece-wise fit is the
  short-dashed line, the \citet{col01} form is the long-dashed
  line]{li08} and \citet[][dot-dashed line]{yuk08}.}
\label{fig:rates_lit}
\end{figure}

In the past decade, volumetric SN~Ia rates have been measured to
varying degrees of accuracy out to redshifts of $z\sim 1.6$
(Fig.~\ref{fig:rates_lit}).  \citet{cap99} compute the SN~Ia rate in
the local universe ($z\sim 0.01$) from a combined visual and
photographic sample of $\sim 10^4$ galaxies, yet their ability to
distinguish core-collapse SNe from Type Ia SNe was severely limited.
More recent work by \citet{li11b} using $\sim270$ SNe Ia from the Lick
Observatory Supernova Search \citep[LOSS;][]{lea11} has made
significant improvements in the statistics over previous studies on
local SNe~Ia.  The rates published by \citet{dil10} include data from
516 SNe~Ia at redshifts $z<0.3$ from the SDSS-II Supernova Survey
(SDSS-SN), with roughly half of these confirmed through spectroscopy.

At intermediate redshifts, rate measurements are provided by
\citet[][38 SNe from the Supernova Cosmology Project in the range
$0.25\leq z \leq 0.85$]{pai02}, \citet[][8 SNe within
$0.3<z<1.2$]{ton03}, and \citet[][$>100$ SNe from the IfA Deep Survey,
23 of which have spectra]{rod10}. \citet{nei06} used a spectroscopic
sample of 58 SNe Ia from the first two years of SNLS to measure a
cumulative volumetric rate in the redshift range $0.2<z<0.6$.

SN~Ia rates out to $z\sim1.6$ are presented by \citet{dah04} using 25
SNe Ia (19 with spectra) from \textit{Hubble Space Telescope (HST)}
observations of the Great Observatories Origins Deep Survey (GOODS)
fields. These data were reanalyzed by \citet{kuz08} using a Bayesian
identification algorithm, and the \textit{HST} sample updated by
\citet{dah08} extending the 2004 sample to 56 SNe. Ground-based
measurements from the Subaru Deep Field have also been made by
\citet{poz07} using 22 SNe~Ia, updated by \citet{gra11} with 150
events.

The general trend of Fig.~\ref{fig:rates_lit} reveals that the rates
typically increase from $z=0$ to $z=1$.  There is a wide spread in the
existing rate measurements, particularly in the range $0.4 < z < 0.8$.
At higher redshifts, data from the GOODS collaboration provide some
apparent evidence for a turnover in the SN~Ia rates.  In particular,
\citet{dah04,dah08} report a decline in SN~Ia rates beyond $z\sim
0.8$.  If present, this decline might point to a larger characteristic
delay time between star formation and SN explosion \citep[see
also][]{str04}.  However, another independent analysis of the
\textit{HST} GOODS data finds rates that are offset, with measurements
by \citet{kuz08} consistently lower than those of \citet{dah04,dah08}.
\citet{kuz08} argue that their results do not distinguish between a
flat or peaked rate evolution.  Ground-based data in this range
\citep{gra11}, while consistent with the \textit{HST}-based results,
show no obvious evidence for a decline above $z\sim1$.

In this paper we use four years of data from the SNLS sample to
investigate the evolution of SN~Ia rates with redshift out to $z\sim
1.1$.  The sample presented comprises $\sim 700$
photometrically identified SNe~Ia from SNLS detected with the
real-time analysis pipeline \citep{per10}. One third of these have
been typed spectroscopically, and one half of the $\sim700$ have a
spectroscopic redshift (sometimes from ancillary redshift surveys in
the SNLS fields).  No other data set currently provides such a
well-observed and homogeneous sample over this range in redshift.

Additionally, rigorous computation of the survey detection
efficiencies and enhancements in photometric classification techniques
are incorporated into the new SNLS rate measurements.  Monte Carlo
simulations of artificial SNe~Ia with a range of intrinsic parameters
are performed on all of the detection images used in the SNLS
real-time discovery \citep{per10}; these provide an exhaustive
collection of recovery statistics, thereby helping to minimize the
effects of systematic errors in the rate measurements.

The SNLS SNe~Ia can be used to examine the relationship between the
\snr\ and redshift, given some model of the SN Ia DTD. The size of the
SNLS sample also permits a division of the SNe Ia by light-curve width
\citep[in particular the ``stretch''; see][]{per97}, allowing a search
for differences in the volumetric rate evolution expected by any
changing demographic in the SN Ia population.  Brighter, more
slowly-declining (i.e., higher stretch) SNe~Ia are more frequently
found in star-forming spirals, whereas fainter, faster-declining
SNe~Ia tend to occur in older stellar populations with little or no
star formation \citep{ham95,sul06b}.  If the delay time for the
formation of the lowest-stretch SNe~Ia is sufficiently long (i.e.,
their progenitors are low-mass stars $\sim 10$ Gyr old), these SNe~Ia
will not occur at high redshifts \citep{how01}.  The behavior of the
high-$z$ rates can reveal the properties of the progenitor systems.

The organization of this paper is as follows: An overview of the rate
calculation is provided in \S\ref{sec:ratecalc}.  The SNLS data set,
along with the light-curve fitting and selection cuts used to define
the photometric sample, is introduced in \S\ref{sec:SNLS}.  SN~Ia
detection efficiencies and the rate measurements are presented in
\S\ref{sec:effs} and \S\ref{sec:SNLSrates}, respectively.  Several
models of the SN Ia DTD are then fit to the rate evolution in
\S\ref{sec:dtds}, and the results discussed.  Finally, the stretch
dependence of the rate evolution is investigated in
\S\ref{sec:stretchdep}. We adopt a flat cosmology with
($\Omega_M$,$\Omega_\Lambda$)=(0.27,0.73) and a Hubble constant of
$H_0=70\,\mathrm{km}\,\mathrm{s}^{-1}\,\mathrm{Mpc}^{-1}$.

\section{The rate calculation}
\label{sec:ratecalc}

The volumetric SN Ia rate in a redshift ($z$) bin $z_1 < z < z_2$ is
calculated by summing the inverse of the detection efficiencies,
$\varepsilon_i$, for each of the $N$ SNe~Ia in that bin, and dividing
by the co-moving volume ($V$) appropriate for that bin
\begin{equation}
\label{eq:rate}
  r_{\mathrm{v}}(z)=\frac{1}{V}\sum_{i=1}^N
  \frac{(1+z_i)}{\varepsilon_i(z_i,s_i,c_i)\,\Delta T_i}.
\end{equation}
The factor $(1+z_i)$ corrects for time dilation (i.e., it converts to a
rest-frame rate), $\Delta T_i$ is the effective search duration in
years, and the volume $V$ is given by
\begin{equation}
\label{eq:volume}
V=\frac{4\pi}{3} \frac{\Theta}{41253}\left[
  \frac{c}{H_0}\int_{z_1}^{z_2}\frac{dz^\prime}{\sqrt{\Omega_M(1+z^\prime)^3+\Omega_\Lambda}}
  \right]^3 \mathrm{Mpc}^3
\end{equation}
where $\Theta$ is the area of a search field in deg$^2$ and in this
equation $c$ is the speed of light, and $H_0$, $\Omega_M$ and
$\Omega_\Lambda$ are the cosmological parameters, and we assume a flat
universe. 

$\varepsilon_i$ is a recovery statistic which describes how each SN Ia
event should be weighted relative to the whole population;
$1-\varepsilon_i$ gives the fraction of similar SNe Ia that exploded
during the search interval but that were not detected, for example due
to sampling or search inefficiencies. $\varepsilon_i$ is a function of
the SN stretch $s$, a dimensionless number that expands or contracts
a template light curve defined as $s=1$ to match a given SN event, the
SN color $c$, defined as the rest-frame $B-V$ color at maximum light
in the rest-frame $B$-band, and the SN $z$. 

The $\varepsilon_i$ are evaluated separately for each year and field
of the survey, and are further multiplied by the sampling time
available for finding each object ($\Delta T_i$) to convert to a ``per
year'' rate. Typically these are 5 months for the SNLS, though this is
dependent on the field and year of the survey. Thus, in practice,
eqn.~(\ref{eq:rate}) is evaluated for each search field and year that
the survey operates.

This ``efficiency'' method is particularly suited for use with
Monte-Carlo simulations of a large, well-controlled survey such as
SNLS.  Its disadvantage is that it is not straight forward to correct
for the likely presence of SNe that are not represented (in $z/s/c$
parameter space) among the $N$ in eqn.~(\ref{eq:rate}) (for example,
very faint or very red SNe Ia) without resorting to assuming a
luminosity function to give the relative fractions of SNe with
different properties. In particular, we are not sensitive to, and nor
do we correct for, spectroscopically peculiar SNe Ia in the SN2002cx
class \citep[e.g.,][]{li03}, and similar events such as SN2008ha
\citep[e.g.,][]{fol08}, super-Chandrasekhar events
\citep[e.g.,][]{how06}, and other extremely rare oddballs
\cite[e.g.,][]{kri11}. We also exclude sub-luminous SNe Ia (here
defined as $s<0.7$, a definition that would include SN1991bg-like
events) but note that these are studied in considerable detail for the
SNLS sample in our companion paper, \citet{gon11}. Thus, we are
presenting a measurement of the rates of ``normal'', low to moderate
extinction SNe Ia (explicitly, $c<0.6$), restricting ourselves to the
bulk of the SN Ia population that we can accurately model. We allow
for these incompletenesses when comparing to other measurements of the
SN Ia rate in \S\ref{sec:dtds}, which do include some of these classes
of SNe Ia.

The photometric sample begins with the set of all possible detections,
to which we apply a series of conservative cuts to remove interlopers.
The SNLS sample and the culling process are described next in
\S\ref{sec:SNLS}.  To each resulting SN~Ia must then be applied the
corresponding $\varepsilon_i$; these are calculated using a detailed
set of Monte Carlo simulations on the SNLS images, a procedure
described in \S\ref{sec:effs}.  The rate results and the measurement
of their associated errors are presented afterwards in
\S\ref{sec:SNLSrates}.

\section{Defining the SNLS sample}
\label{sec:SNLS}

In this section, we describe the SNLS search and the SN Ia sample that
we will subsequently use for our rate analysis. The SNLS is a rolling
SN search that repeatedly targeted four $1\degr\times 1\degr$ fields
(named D1--4) in four filters ($g_Mr_Mi_Mz_M$) using the MegaCam
camera \citep{bou03} on the 3.6~m Canada--France--Hawaii Telescope
(CFHT).  The SNLS benefited from a multi-year commitment of observing
time as part of the CFHT Legacy Survey.  Queued-service observations
were typically spaced $3-4$ days apart during dark/grey time, yielding
$\sim 5$ epochs on the sky per lunation. Key elements of the SNLS are
its consistent and well-defined survey characteristics, and the
high-quality follow-up spectroscopy from 8m-class telescopes such as
Gemini \citep{how05,bro08,wal11}, the ESO Very Large Telescope
\citep[VLT;][]{bal09}, and Keck \citep{ell08}. Due to the finite
amount of follow-up observing time available, not all of the SN Ia
candidates found by SNLS were allocated for spectroscopic follow-up
\citep[for a description of follow-up prioritization,
see][]{sul06a,per10}.  The availability of well-sampled light curves
and color information from the SNLS nonetheless allow us to perform
photometric identification and redshift measurements, even in the
absence of spectroscopic data.

\begin{deluxetable}{crrcc}
\tablewidth{0pt}
\tablecaption{SNLS fields and survey parameters.\label{tab:deepfields}}
\tablehead{
  \colhead{Field} & \colhead{RA (J2000)} & \colhead{DEC (J2000)} &
    \colhead{Area (sq.\ deg)} & \colhead{N$_{\mathrm{seasons}}$}}
\startdata
D1 & 02:26:00.00 & -04:30:00.0 & 0.8822 & 4\\
D2 & 10:00:28.60 & +02:12:21.0 & 0.9005 & 4\\
D3 & 14:19:29.01 & +52:40:41.0 & 0.8946 & 4\\
D4 & 22:15:31.67 & -17:44:05.7 & 0.8802 & 4\\
\enddata
\end{deluxetable}

To identify the photometric SN~Ia sample, we begin with all variable
object detections in the SNLS real-time
pipeline\footnote{http://legacy.astro.utoronto.ca} \citep{per10}.
Other articles will describe a complementary effort to measure the
rates with a re-analysis of all of the SNLS imaging data
\citep[e.g.,][]{baz11}.  We use SNLS data up to and including the
fifth year of D3 observing in June, 2007\footnote{In June 2007, the
  $i_M$ filter on MegaCam was damaged during a malfunction of the
  filter jukebox.  Candidates discovered after this period were
  observed with a new $i_M$ filter, requiring new calibrations for
  subsequent images, and were thus not included in the present
  study.}. The first (2003) season of D3 is omitted in this analysis;
this was a pre-survey phase when the completeness of the SN data
differed significantly from the rest of the survey.  The remaining
detections made during four observing seasons for each of the four
deep fields are considered in this analysis.  Each period of
observation on a given field is called a ``field-season'', with 16
field-seasons in total (4 fields observed for 4 seasons).  The
coordinates of the field centers and other information are provided in
Table~\ref{tab:deepfields}.

We remove all candidates falling within masked regions in the deep
stacks.  These regions include areas in and around saturated bright
stars or galaxies, as well as in the lower signal-to-noise edge
regions of the dithered mosaic.  The remaining unmasked areas in each
field are listed in Table~\ref{tab:deepfields}, and add up to a total
of $3.56$ square degrees.  Galaxy catalogs from these image stacks are
used to determine the placement of test objects in the simulations
described later in \S\ref{sec:effs}.  This cut therefore ensures that
the areas being considered in the rates calculation match those used
in the detection efficiency measurements.

We next fit each event with a light curve fitter to determine its
redshift (where no spectroscopic redshift is available) and
photometric parameters ($\S$\ref{sec:light-curve-fitting}). We then
remove SN Ia candidates with insufficient light curve coverage
(\S\ref{sec:culls}). Finally, we use the light curve fits to identify
and remove core-collapse SNe as well as other transients, such as AGN
and variable stars (\S\ref{sec:removing-non-sne}). Each of the
remaining SNe~Ia will then correspond to some fraction of the true
number of events having similar photometric properties but which were
undetected by our survey. This detection efficiency will be determined
from the Monte Carlo simulations presented in \S\ref{sec:effs}.

\subsection{Light-curve fitting}
\label{sec:light-curve-fitting}

We fit template light curves to the SN Ia candidates to identify those
that don't match typical SNe~Ia.  Flux measurements are made on all of
the final ``Elixir-preprocessed'' images\footnote{CFHT-LS images
  processed with the Elixir pipeline are available from the Canadian
  Astronomy Data Centre (CADC):
  http://www.cadc-ccda.hia-iha.nrc-cnrc.gc.ca/cadc/} \citep{mag04}.
The Canadian SNLS photometric pipeline \citep{per10} was used to
measure the SN fluxes, using images processed with the accumulated
flat-fields and fringe-maps from each queue run, and aligning
photometrically to the tertiary standard stars of \citet{reg09}.

Two light-curve fitting tools were used to help identify the SNe Ia:
{\tt estimate\_sn} \citep{sul06a} for preliminary identification and
for measuring SN Ia photometric redshifts, and SiFTO \citep{con08} for
final light-curve fitting to measure the stretch and color of each
candidate.  The {\tt estimate\_sn} routine is not designed for exact
measurement of SN Ia parameters, and SiFTO does not fit for redshift, so
we require this two-step process to fully characterize the photometric
sample of events.

In {\tt estimate\_sn}, the measured fluxes in $g_Mr_Mi_Mz_M$ are fit
using SN~Ia spectral templates from \citet{hsi07}.  The current
version of the code includes the addition of priors in stretch, color,
and $\Delta$mag.  These are determined from the distributions measured
for the spectroscopic sample.  The photometric redshifts
($z_{\mathrm{SNphot}}$) output from this routine are used for
candidates with no available spectroscopic redshifts from either the
SN or its host.  SiFTO is an empirical light-curve analysis method
that works by manipulating a spectral energy distribution (SED) model
rather than operating in integrated filter space \citep{con08}.  SiFTO
does not impose a color model to relate the observed filters during
the light-curve fits.  The implication of this is that SiFTO cannot
easily be modified to fit for redshift, and thus requires a known
input $z$. Output SN Ia fits are parameterized by stretch, date of
maximum light, and peak flux in each filter.

\begin{figure}
\includegraphics[width=0.8\textwidth]{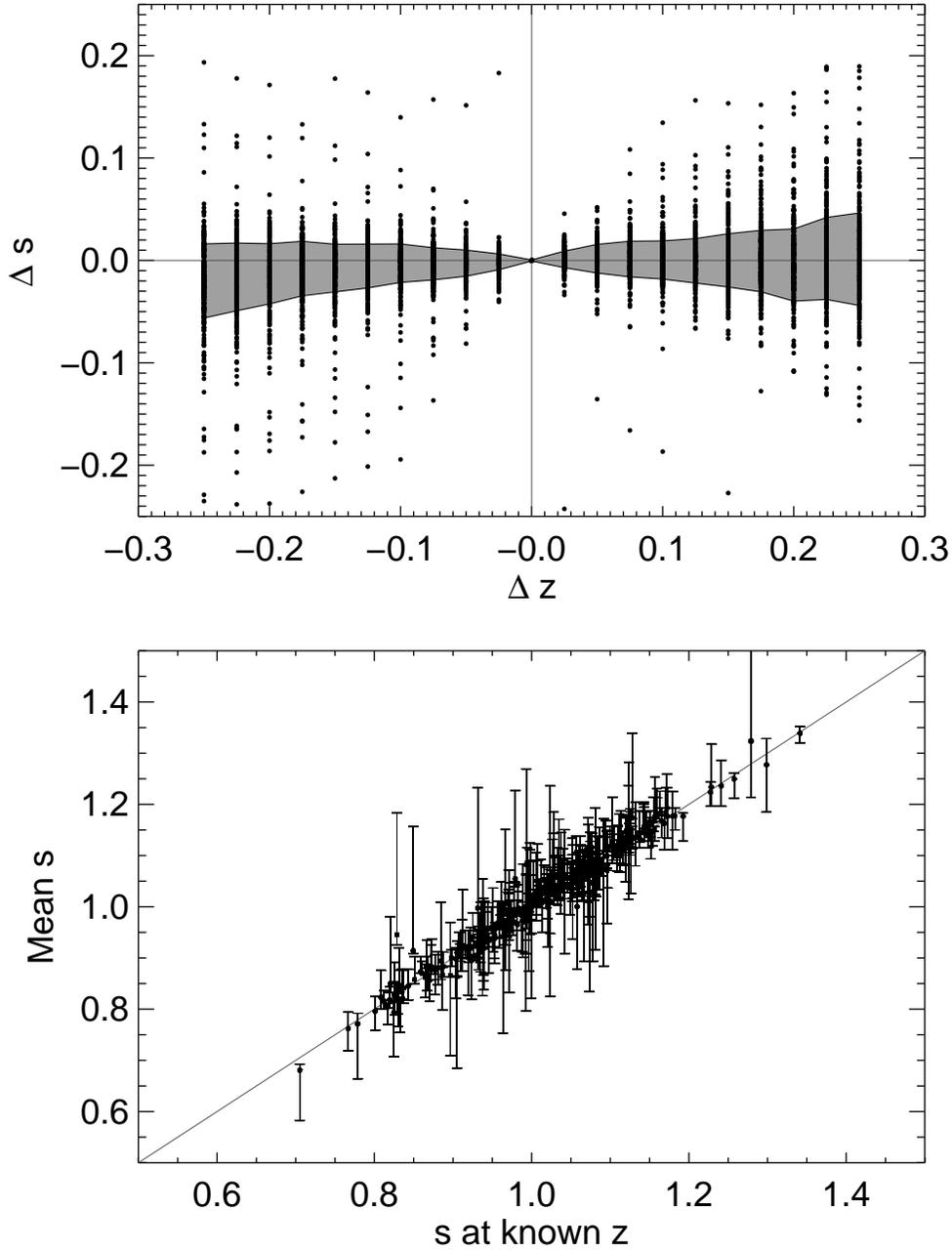}
\caption{The effects of deliberately shifting the input redshift to
  SiFTO. The top plot shows the change in output stretch for confirmed
  SNe~Ia from the SNLS sample as the input redshift for the SiFTO fit
  is offset from $z_{\mathrm{spec}}-0.3$ to $z_{\mathrm{spec}}+0.3$.
  The gray shaded area represents the standard deviation of the
  measured points about the median $\Delta s$.  The bottom plot shows
  the mean output stretch for each SN~Ia as a function of its known
  stretch (at zero redshift offset).  The error bars for each SN~Ia in
  the lower plot represent the full range in stretch values output
  from SiFTO as the input redshift is changed.}
\label{fig:sinvar}
\end{figure}

The stretch measurement provided by SiFTO is largely invariant to
changes in input redshift, as demonstrated in Fig.~\ref{fig:sinvar}.
Even when the input redshift is off by $\Delta z = \pm 0.3$, the
output stretch remains within $s\pm 5\%$ of its actual value.  Opacity
effects in the SN ejecta are more pronounced in the bluer bands,
causing a more rapid decline \citep{kas07}; light curves measured at
shorter wavelengths are therefore intrinsically narrower.  As a
result, if SiFTO is (for example) given an incorrectly small input
redshift, it measures the ``wrong'' time dilation but simultaneously
samples further towards the blue end of the spectrum where the
template is intrinsically narrower.  The latter effect partially
negates the first, resulting in the same stretch measurement
regardless of marginal deviations in input redshift.  While this is
not similarly true of the derived color or measured fit quality, this
stretch invariance is extremely useful for establishing an initial
constraint in fitting photometric redshifts with {\tt estimate\_sn}.

Spectroscopic redshifts are available for 525 (43\%) of the detections
remaining after the observational cuts: 420 from SNLS spectroscopy and
the rest from host-galaxy measurements (including data from DEEP/DEEP2
\citep{dav03}, VVDS \citep{lef05}, zCOSMOS \citep{lil07}, and
additional SNLS VLT MOS observations). The external redshifts are
assigned based on a simple RA/DEC matching between the SNLS and the
redshift catalogues, with a maximum allowed separation of 1.5\arcsec.
For the SNLS MOS work, the host was identified following the
techinques of \citet{sul06b}. The known redshifts are then held fixed
in the light-curve fits. We also considered the use of galaxy
photometric redshifts for the SNLS fields \citep[e.g.][]{Ilb06}.
However, though these catalogs have an impressive precision, they tend
to be incomplete and untested below a certain galaxy magnitude. SN Ia
photometric redshifts do not suffer these problems.

\begin{figure}
\plotone{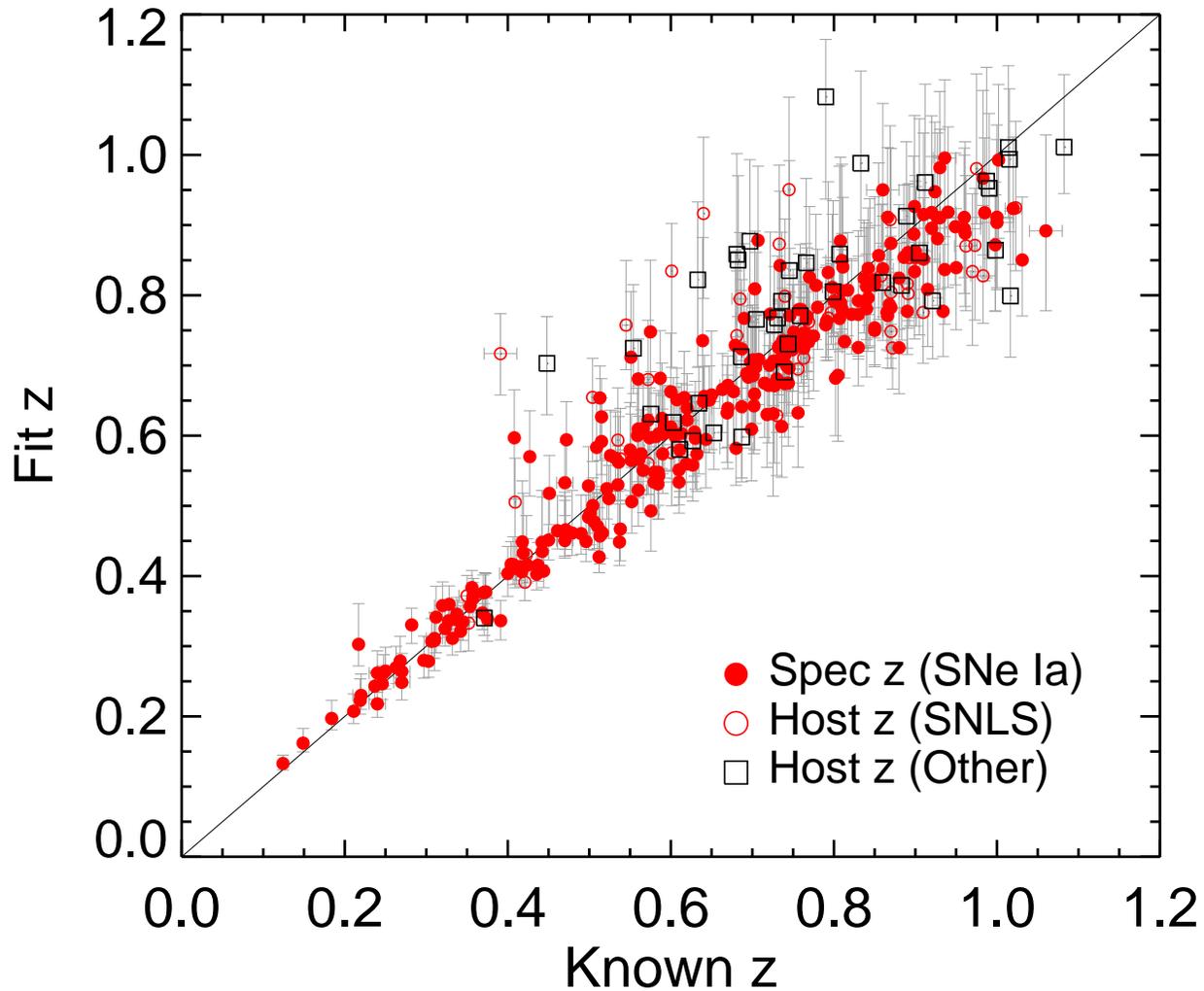}
\caption{The comparison of SN photo-$z$ measurements to spectroscopic
  redshifts for candidates in the SNLS sample.  Filled (open) circles
  represent confirmed (unconfirmed) SNe~Ia with spectroscopic
  redshifts from the SNLS, and open squares are candidates with host
  spectroscopic redshifts from the literature.}
\label{fig:zcomp}
\end{figure}

SN photometric redshifts (photo-$z$'s) are calculated for the
remaining objects using a multi-step procedure.  Preliminary redshift
estimates are obtained using a first round of {\tt estimate\_sn} fits
without any constraints on the input parameters.  The resulting fit
redshifts are then used as input to SiFTO to measure the stretch for
each object.  These stretch values are then fixed in a subsequent
round of {\tt estimate\_sn} fits to obtain a more robust measurement
of the SN redshift -- constraining at least one input parameter to
{\tt estimate\_sn} improves the quality of the light-curve fits.
Fig.~\ref{fig:zcomp} shows that the $z_{\mathrm{SNphot}}$ are in good
agreement with the spectroscopic redshifts ($z_{\mathrm{spec}}$) out
to $z\ga 0.7$, with a small systematic offset above that.  The median
precision in $z_{\mathrm{SNphot}}$ for the confirmed SNe~Ia is
\begin{displaymath}
\mathrm{MEDIAN}\left(\frac{|\Delta z|}{(1+z_{\mathrm{spec}})}\right)=0.019
\end{displaymath}
with $\sigma_{|\Delta z|/(1+z_{\mathrm{spec}})} = 0.031$.  For
comparison, \citet{sul06a} find $|\Delta
z|/(1+z_{\mathrm{spec}})=0.031$ with a smaller sample and real-time
data (and a previous version of the {\tt estimate\_sn} code).  In
\S\ref{sec:errsims}, we describe how these $z_{\mathrm{SNphot}}$
errors and the systematic offset are incorporated into the rates
analysis.

\subsection{Light curve coverage cuts}
\label{sec:culls}

Each candidate must pass a set of light curve quality checks to be
included in the photometric sample of SNe~Ia for the rate calculation.
Requiring that the SN light curves are well measured ensures that the
photometric typing technique is more reliable, and that it is straight
forward to correct for the effects of the selection cuts on the rates
themselves. Therefore, candidates with insufficient light-curve
coverage to measure accurate redshift, stretch, and color values from
template fits are removed from the detected sample.  We define
observational criteria in terms of the phase, $t$, of the SN in
effective days (d) relative to maximum light in the rest-frame
$B$-band, where
\begin{equation}
  t_{\mathrm{eff}} = \frac{t_{\mathrm{obs}}}{s(1+z)},
\end{equation}
and $t_{obs}$ is the observer-frame phase of the SN. The time of
maximum light is determined using the light curve fitter SiFTO
\citep{con08}, described in the previous section.

Each object is required to have a minimum of each of the following:
\begin{itemize}
\item One observation in each of $i_M$ and $r_M$ between $-15$d and
  $+2.5$d for early light-curve coverage and color information;
\item One observation in $g_M$ between $-15$d and $+5$d for additional
  color information;
\item One observation in each of $i_M$ and $r_M$ between $-9$d and
  $+7$d for coverage near peak;
\item One observation in either $i_M$ or $r_M$ between
$+5$d and $+20$d to constrain the later stages of the light curve.
\end{itemize}
These conditions differ slightly from those used by \citet{nei06} in
their analysis of the first year of SNLS data.  Note that no cuts are
made on the signal-to-noise (S/N) on a particular epoch; that is, a
detection of a candidate on each of the observation epochs is not a
requirement. We also neglect the redshift offset seen in
Fig.~\ref{fig:zcomp} in calculating the above rest-frame epochs. We
estimate that this would shift the effective epochs by only one day in
the worst case ($+20$d; a $z=1$ SN), and in most cases would be far
smaller than this.

Table~\ref{tab:culls} provides the numbers of candidates that survive
each of the applied cuts.  In total, $1210$ SNLS detections pass the
light curve coverage cuts, $305$ of which are spectroscopically
confirmed SNe~Ia.  (For consistency, these same objective requirements
are also applied to the artificial SNe~Ia used in the Monte Carlo
simulations (\S\ref{sec:effs}), thereby directly incorporating the
effects of this cut into the detection efficiencies.)  With these
objective criteria satisfied, we can then use light-curve fitting to
define a photometric SN~Ia sample.

\begin{deluxetable}{lcc}
\tablewidth{0pt}
\tablecaption{Number of candidates after selection cuts\label{tab:culls}}
\tablehead{
  \colhead{Cut} & \colhead{All candidates} & \colhead{Confirmed SNe~Ia}
}
\startdata
Masking cut              & 1538 & 325 \\
Observational cuts       & 1210 & 305 \\
Fit quality and $s$ cuts & ~691 & ~286
\enddata
\end{deluxetable}

\subsection{Removing non SNe Ia}
\label{sec:removing-non-sne}

A set of $\chi^2_{\nu}$ goodness-of-fit cuts (here, $\chi^2_{\nu}$ is
the $\chi^2$ per degree of freedom, $\nu$) are applied to all of the
SN~Ia light-curve fits from {\tt estimate\_sn} to help eliminate
non-Ias from the current sample \citep[see also][]{sul06a}.  An
overall $\chi^2_{\nu}$ cut along with individual $i_M$ and $r_M$
filter $\chi^2_{\nu}$ constraints are applied separately for cases
both with and without known redshifts.  Light-curve fit quality limits
for the sample with known input redshifts are set to $\chi^2_{\nu}<9$,
$\chi_i^2<9$, and $\chi_r^2<18$ (here the $\nu$ is omitted for
clarity).  Those with fit redshifts are given stricter limits of
$\chi^2_{\nu}<6$, $\chi_i^2<6$, and $\chi_r^2<12$.  The tighter
$\chi^2_{\nu}$ limits for candidates without known redshifts are necessary
since core-collapse SNe --- specifically SNe Ib/c --- can sometimes
achieve better fits to SN~Ia templates when $z$ is permitted to float
from the true value.  The limits are determined empirically by
maximizing the fraction of SNe~Ia remaining in the sample, while also
maximizing the number of known non-Ias that are removed. Note that
{\tt estimate\_sn}, unlike SiFTO, enforces a color relation between
the fluxes in different filters, which leads to larger $\chi^2_{\nu}$ than
in SiFTO fits.

In the case of fixed [floating] redshifts input to the fit, $>95\%$
[$>96\%$] of spectroscopically identified SNe~Ia survive the
$\chi^2_{\nu}$ cuts (we correct for this slight inefficiency when
calculating our final rate numbers), while $0\%$ [$13\%$] of confirmed
non-Ias remain in the sample.  A final round of SiFTO fits is then
performed to determine the output values of stretch and color.  The
input redshifts to SiFTO are set to $z_{\mathrm{SNphot}}$ wherever no
$z_{\mathrm{spec}}$ values are available.

\begin{figure*}
\includegraphics[width=0.49\textwidth]{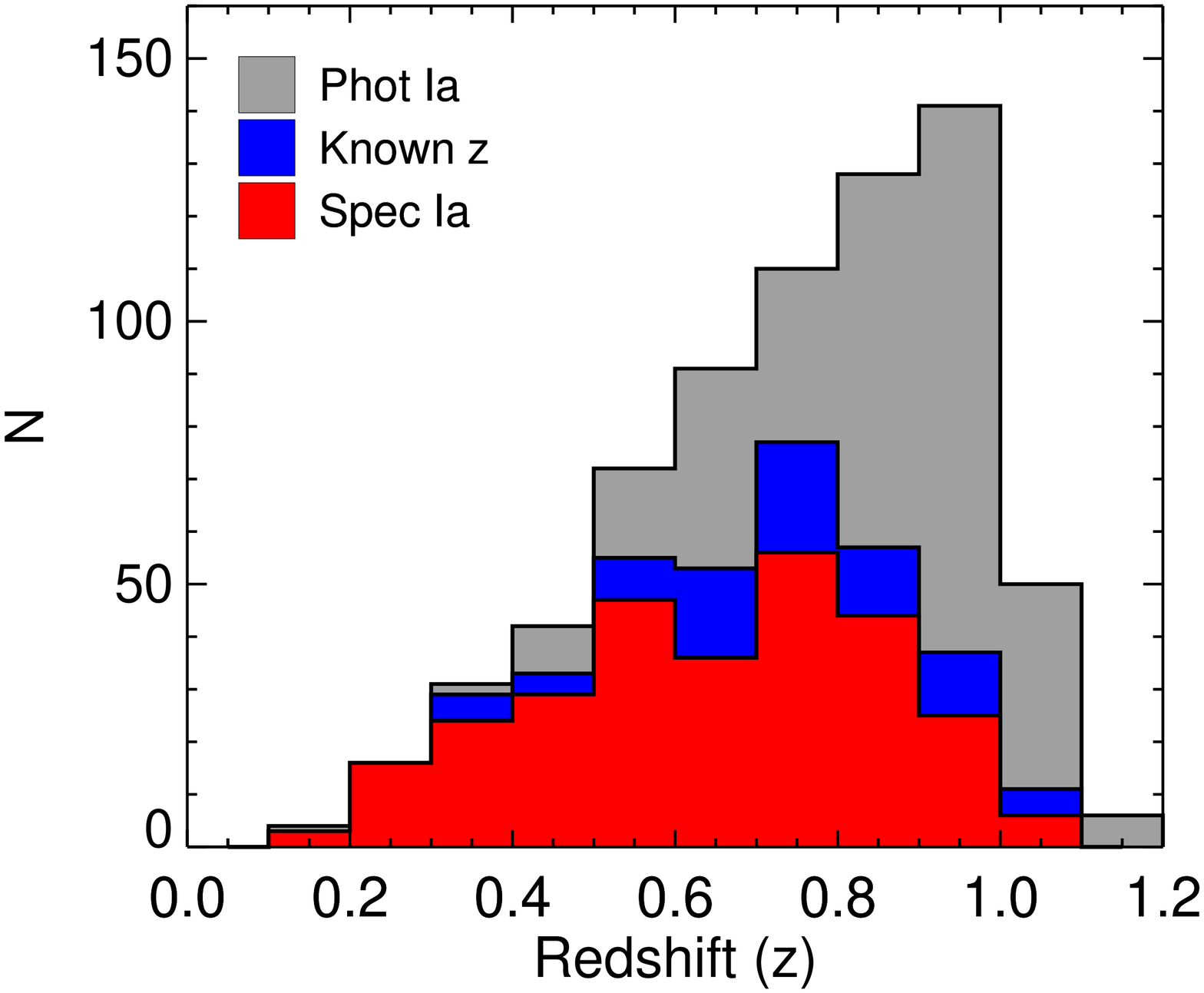}
\includegraphics[width=0.49\textwidth]{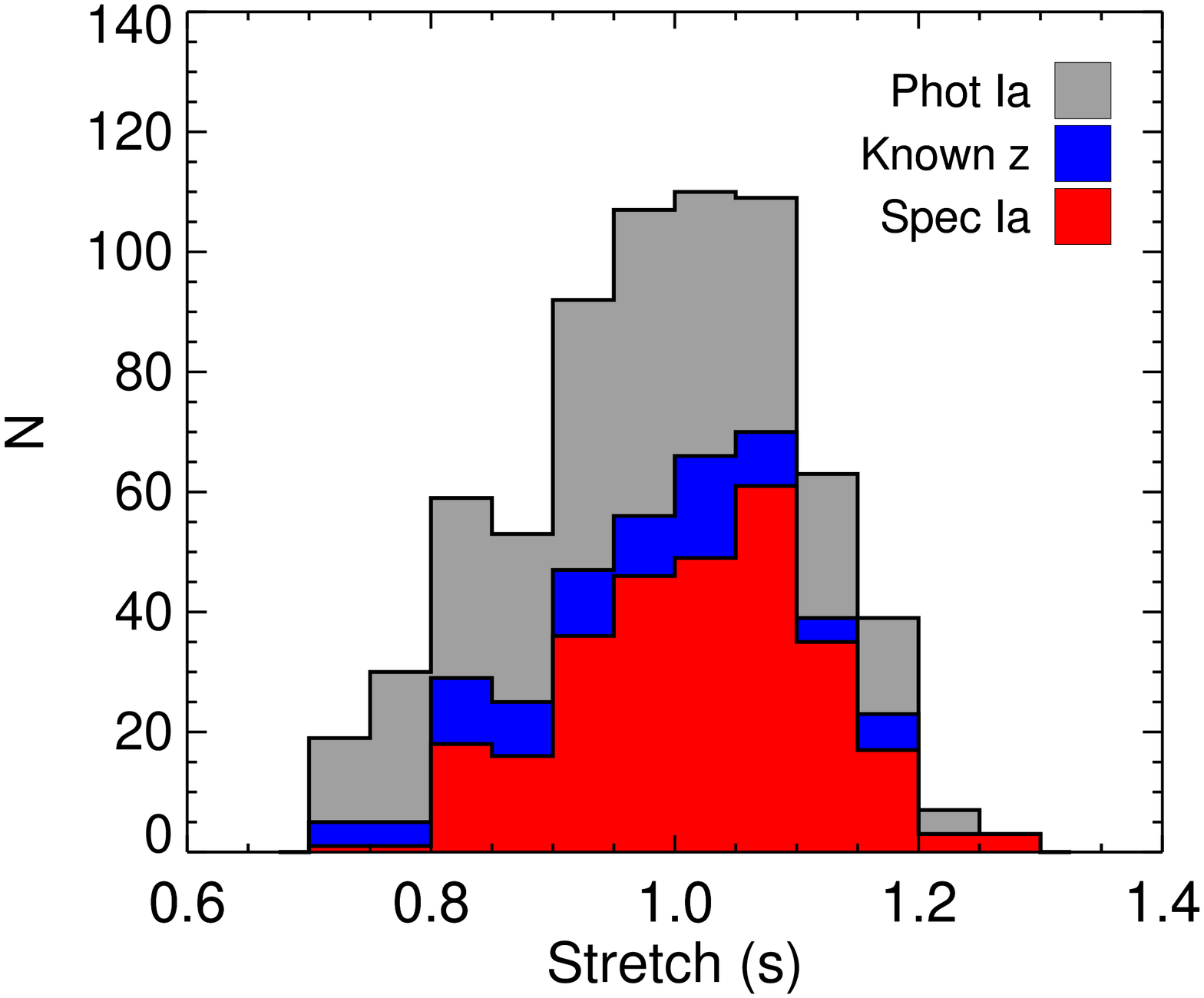}
\includegraphics[width=0.49\textwidth]{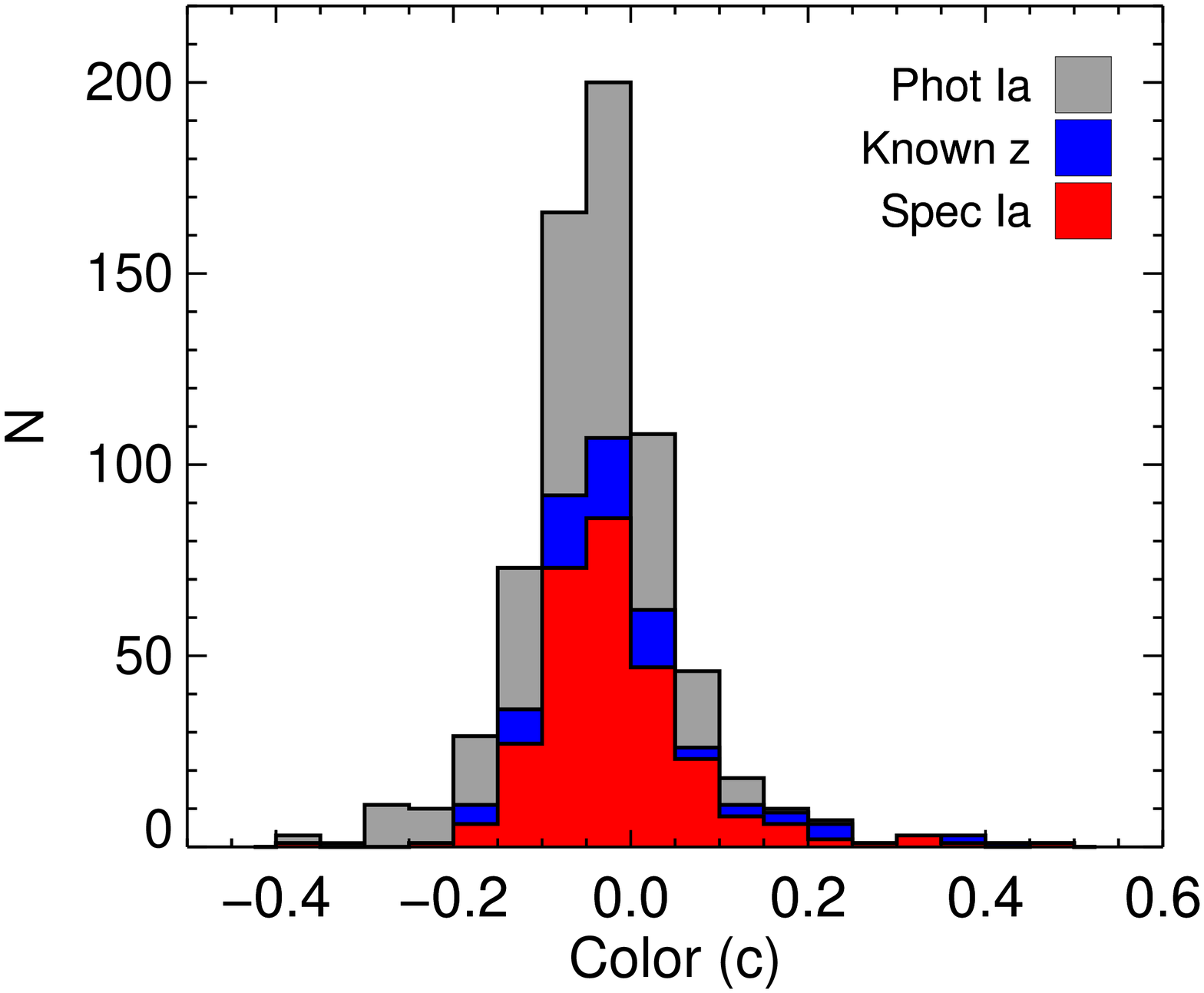}
\caption{Distributions in redshift (upper left), stretch (upper
  right), and color (lower left) for the SNLS SNe~Ia.  The gray
  histogram represents the final photometric SN Ia sample and the blue
  histogram shows the fraction of the sample with known redshifts. The
  spectroscopically confirmed SNe~Ia are shown as the red distribution
  in each plot.  Sample incompleteness causes the decline in the
  observed population at $z>1.0$.}
\label{fig:objpars}
\end{figure*}

One final light-curve fitting cut is then applied on the sample,
requiring that the output SiFTO template fits have
$\chi^2_{\mathrm{SiFTO}}<4$.  This step removes all but one of the
remaining confirmed non-Ias\footnote{The identification of
  SNLS-06D4cb is inconclusive, although it has a spectrum that is a
  poor match to a SN Ia.  The SN photometric redshift for this object
  is $z_{fit}=0.64$ but the host has a spectroscopic redshift of
  $z=0.4397$.} when all redshifts are allowed to float, while at the
same time maximizing the number of confirmed SNe~Ia passing the cut.
No known contaminants remain when all available $z_{\mathrm{spec}}$
values are fixed in the fits.

\subsection{The photometric SN I\lowercase{a} sample}

The final photometric SN~Ia (phot-Ia) sample is restricted to $0.1\leq
z\leq 1.1$. Above this redshift, the rates are found to be too
uncertain to include in subsequent analyses.  This is a result of low
S/N, poor detection efficiency, 100\% spectroscopic incompleteness,
and the potential for increased contamination from non-Ias. Only
candidates having stretch values within $0.7\leq s\leq 1.3$ are
considered in the present study.  This range is characteristic of the
SNLS spectroscopic sample --- shown by the red histogram in the
central plot of Fig.~\ref{fig:objpars} -- but excludes sub-luminous
events such as SN1991bg. These sub-luminous, low-stretch SNe~Ia in the
SNLS sample have been studied in detail by \citet{gon11} -- our
stretch limit removes 22 such objects from our sample.  Extremely red
($c>0.6$), and presumably highly extincted, candidates are also
removed.  This cut eliminates only one event: SNLS-04D2fm, a faint SN
of unknown type at $z_\mathrm{spec}=0.424$.

The final redshift, stretch, and color distributions resulting from
the various cuts are shown in Fig.~\ref{fig:objpars}.  The phot-Ia
sample consists of $691$ objects, $371$ of which have known redshifts.
A total of $286$ objects in this sample have been spectroscopically
confirmed as Type Ia SNe (Table~\ref{tab:culls}).  The redshift
histogram reveals that the incompleteness of the spectroscopic sample
(in red) begins to increase beyond $z\sim 0.5$, where the rise in
the required exposure time makes taking spectra of every candidate too
expensive.  The effects of incompleteness in the observed SNLS sample
become severe above $z>1.0$. The full phot-Ia sample has median
stretch and color values of $s=1.00$ and $c=-0.04$, respectively.  The
color distribution peaks at a slightly redder value than the estimated
typical color of a SNe~Ia of $c_f \sim -0.06$, based on the
distribution observed for the spectroscopic SNLS sample.

\section{Detection efficiencies}
\label{sec:effs}

With the final SN Ia sample in hand, we now need to estimate the
weight that each of these events contributes in our final rates
calculation. These ``detection efficiencies'' depend on many
observational factors and will obviously vary with SN Ia
characteristics.  For example, at higher-redshift, the higher stretch
SNe Ia are more likely to be recovered not only because they are
brighter, but also because they spend a longer amount of time near
maximum light, and are therefore more likely to pass the culls of
$\S$\ref{sec:culls}.  In a rolling search like SNLS, such effects can
be directly accounted for by measuring recovery statistics for a range
of simulated input SN Ia properties using the actual images (and their
epochs) observed.  This is a brute-force approach, but is a practical
way to accurately model a survey such as SNLS, helping to control
potential systematic errors by avoiding assumptions about image
quality limitations and data coverage that may bias the rate
calculation.  Uncertainties on search time and detection area are
avoided since the actual values are well defined.

\subsection{Monte Carlo simulations}
\label{sec:MC}

An exhaustive set of Monte Carlo simulations were performed for each
field-season to determine the recovery fraction as a function of
redshift, stretch, and color. Full details about these simulations are
presented in \citet{per10}.

A total of $2.5\times 10^6$ artificial SNe~Ia with a flat redshift
distribution were added to galaxies present in the SNLS fields.  Each
host galaxy was chosen to have a photometric redshift within 0.02 of
the artificial SN redshift, with the probability of selecting a
particular galaxy weighted by the ``A+B'' SN rate model with
coefficients from \citet[][hereafter \citetalias{sul06b}]{sul06b}.
Within their host galaxies, the artificial SNe were assigned
galactocentric positions drawn from the two-dimensional Gaussian
distribution about the host centroid returned by SExtractor, i.e., the
artificial SNe are placed with a probability that follows the light of
the host galaxy.

The simulated objects were assigned random values of stretch from a
uniform distribution in the range $0.5\leq s \leq 1.3$, with colors
calculated from the stretch--color relationships presented in
\citet{gon11} (the use of a uniform distribution in stretch ensures
that the parameter space of SN Ia events is equally sampled).  Peak
apparent rest-frame $B$ magnitudes ($m_B$) at each selected redshift
were calculated for our cosmology and a SN Ia absolute magnitude, and
adjusted for the color--luminosity and stretch--luminosity relations.
We use an empirical piece-wise stretch-luminosity relationship with
different slopes above and below $s=0.8$ \citep[e.g.][]{gar04,gon11},
and SN Ia photometric parameters from the SNLS3 analysis
\citep{con11,sul11}. These peak apparent magnitudes were then further
adjusted by an amount $\Delta$mag according to the observed intrinsic
dispersion ($\sigma_\mathrm{int}$) in SN~Ia magnitudes following $s$
and $c$ corrections. Here, $\sigma_\mathrm{int}$ parameterizes a
Gaussian distribution from which a $\Delta$mag can be assigned for
each artificial event.

The SN color--luminosity relation includes both effects intrinsic to
the SN, and extrinsic effects such as dust. We use coefficients
consistent with the SNLS3 analysis, which favor a slope between $m_B$
and $c$ of $<4.1$, the value expected based on Milky Way dust. As
there is no evidence that this slope evolves with redshift
\citep{con11}, we keep it fixed for all the artificial SNe. For the
detection efficiency grids, our $c$ values range up to 0.6,
corresponding to a SN that is $\sim1.8$\,mag fainter in $B$-band than
a normal SN Ia.

\begin{figure}
\plottwo{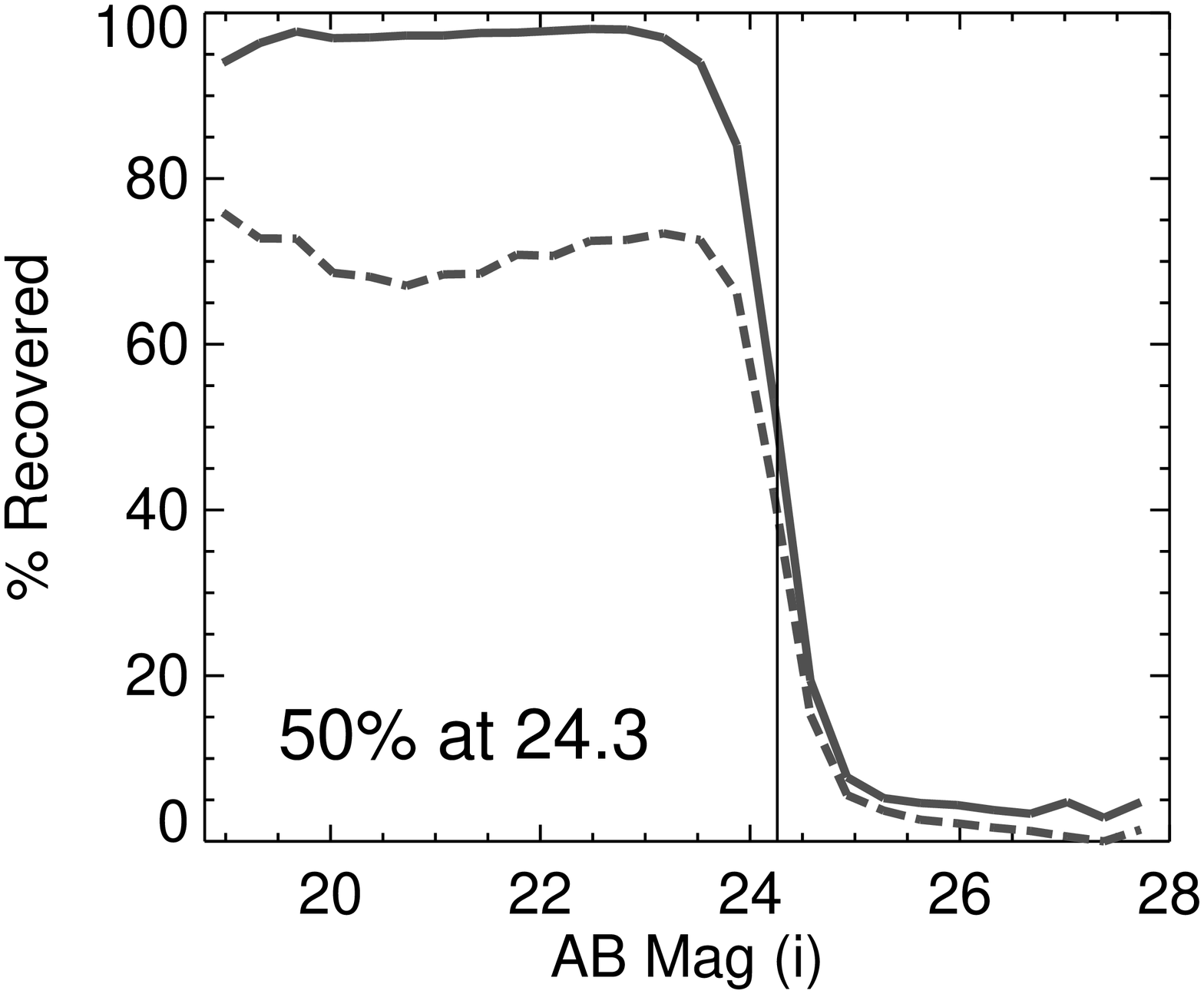}{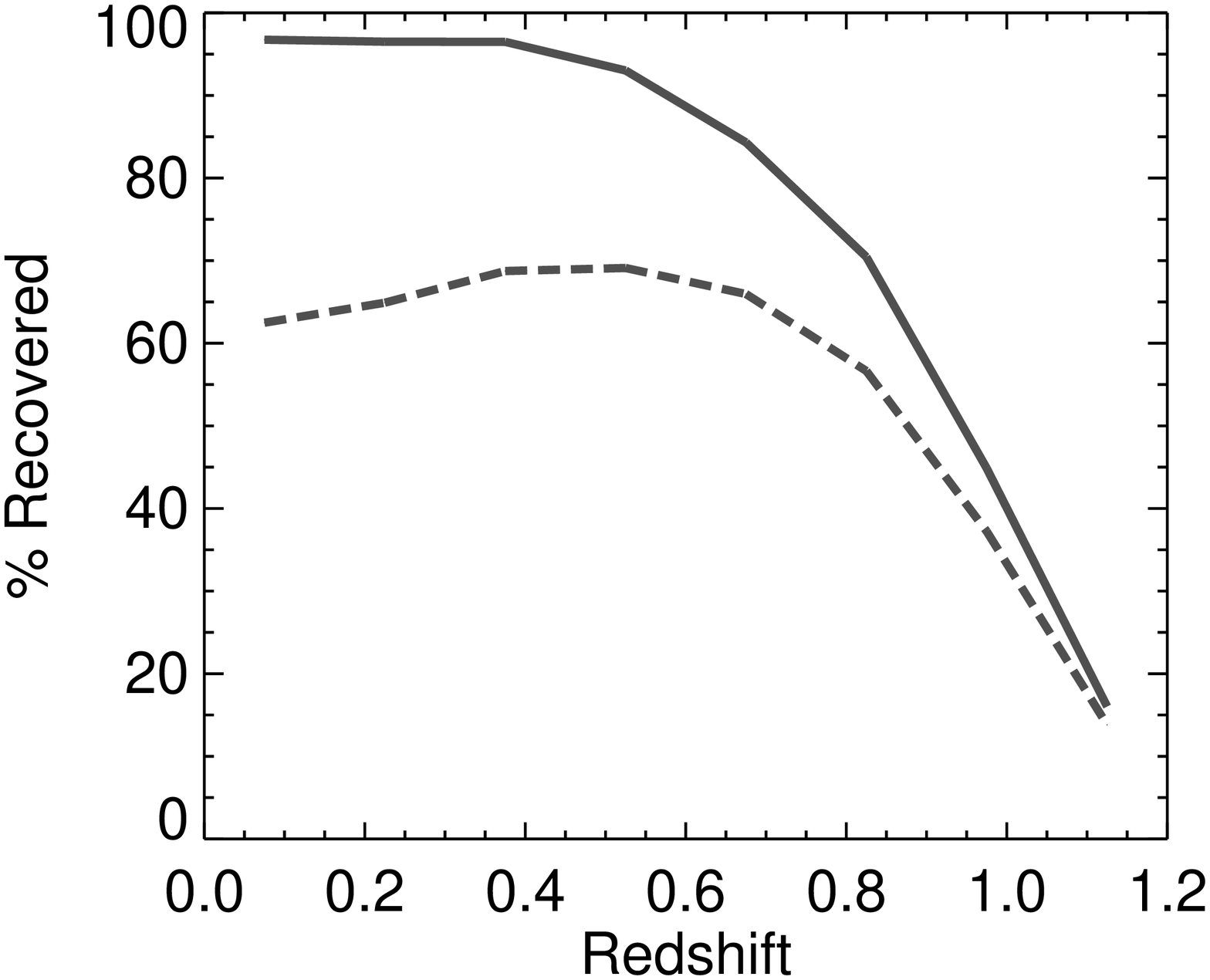}
\plottwo{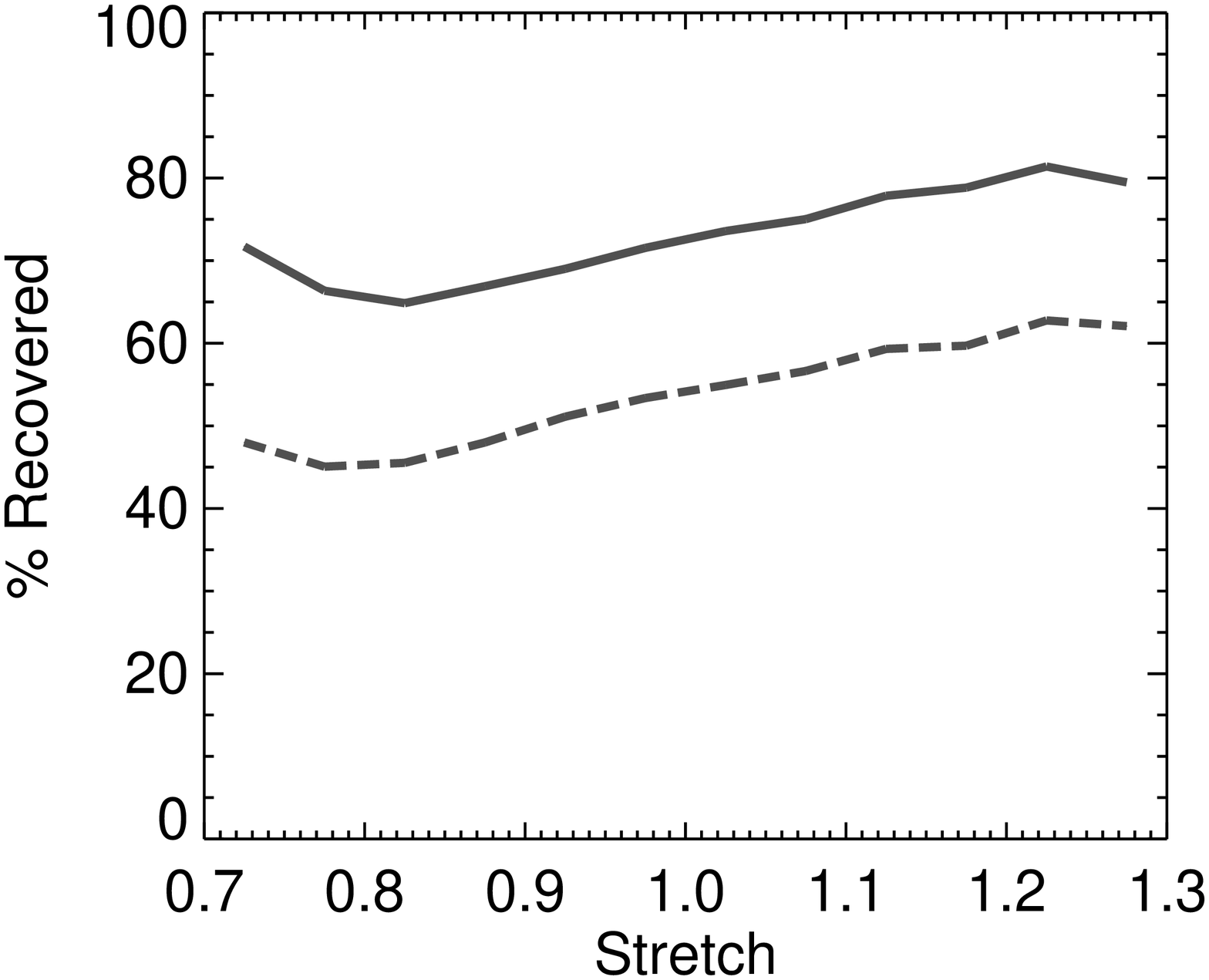}{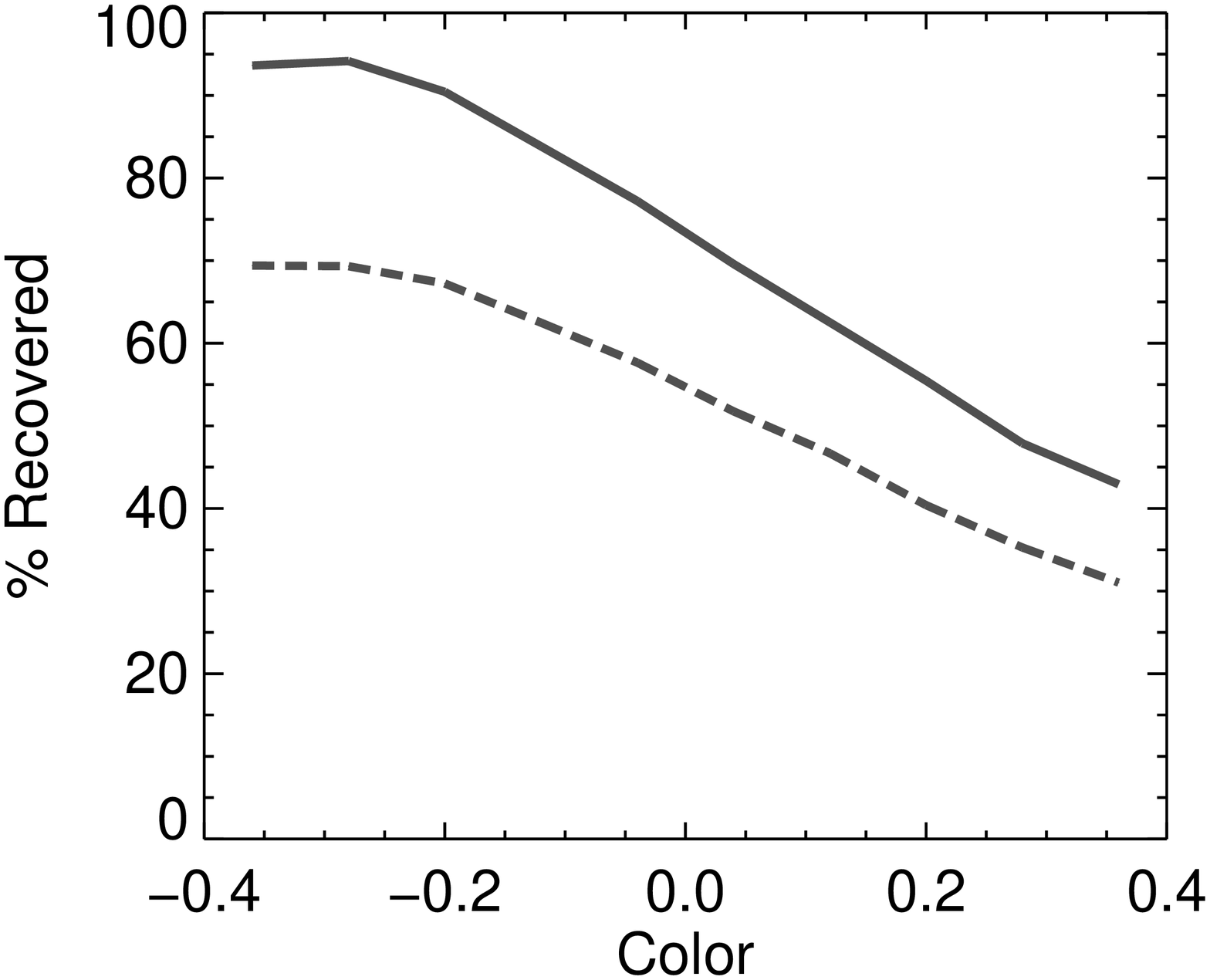}
\caption{Recovery fraction as a function of $i_M$ (AB) magnitude
  (upper left), redshift (upper right), stretch (lower left), and
  color (lower right) for all field-seasons combined.  The solid lines
  represent the fraction of objects found, and the dashed lines
  include the additional observational constraints as described in
  \S\ref{sec:culls}.  These plots include only the artificial SNe~Ia
  from the simulations that lie within the parameter space typical of
  the observed SNLS sample.}
\label{fig:compl}
\end{figure}

Each artificial SN was assigned a random date of peak magnitude. For
the field-season under study, this ranged from 20 observer-frame days
before the first observation, to 10 days after the last observation.
This ensures that the artificial events sample the entire phase range
allowed by the culls in $\S$~\ref{sec:culls} at all redshifts. The
light curve of each event in $i_M$ was then calculated using the
$k$-correction appropriate for each epoch of observation, and each
artificial object was added at the appropriate magnitude into every
$i_M$ image.  The real-time search pipeline \citep{per10}, the same
one that was used to discover the real SNe, was run on each epoch of
data to determine the overall recovery fraction as a function of the
various SN~Ia parameters.  The variation in candidate recovery over
magnitude, redshift, stretch, and color are shown by the solid lines
in Fig.~\ref{fig:compl}.  The 50\% detection incompleteness limit lies
at $i_M=24.3$ mag in the AB system.

As expected, SNe Ia that are high stretch, blue, or at lower redshift
are all generally easier to recover. Note that at lower redshifts, the
faster (less time-dilated) nature of the SN~Ia light curves means that
the observational criteria of \S\ref{sec:culls} are slightly more
likely to remove events (as there are fewer opportunities to observe a
faster SN), hence the observed decrease in the recovered fraction
towards lower-redshifts.  That is, a low-$z$ SN that peaks during
bright time is less likely to be recovered than a higher-$z$ SN
peaking at the same epoch, even if they had the same observed peak
magnitude. This is also partially reflected in the fraction recovered
as a function of magnitude, with a curvature in the recovered fraction
towards brighter magnitudes. The recovery results are discussed in
detail in \citet{per10}.

A grid of detection efficiencies was constructed independently for
each field-season using the recovery statistics in bins of measured
redshift ($\Delta z=0.1$), stretch ($\Delta s=0.1$), and color
($\Delta c=0.2$).  These bin sizes were found to provide adequate
resolution in each parameter. We investigated the use of a higher
resolution in stretch and color, and found no significant on impact
our results. Every observed phot-Ia in the SNLS sample is thereby
assigned a detection efficiency by linearly interpolating in $z/s/c$
space that corresponds to the field-season during which it was
detected, along with its other measured parameters:
$\varepsilon(\mathrm{field},z,s,c)$.  These detection efficiencies are
plotted in Fig.~\ref{fig:objeffs_raw} prior to any adjustments for
sampling time and the availability of observations.  Redder,
lower-stretch SNe~Ia tend to have smaller detection efficiencies, as
shown by the open circles in Fig.~\ref{fig:objeffs_raw}.  For clarity,
detection efficiency errors are not shown in
Fig.~\ref{fig:objeffs_raw}.

Statistical uncertainties on $\varepsilon(\mathrm{field},z,s,c)$ for
well-sampled data are governed by the number of Monte Carlo
simulations performed, and are small in comparison to the systematic
error resulting from assumptions made about the underlying intrinsic
SN~Ia magnitude dispersion (the $\Delta$mag distribution, parameterized
by $\sigma_\mathrm{int}$).  To estimate these latter errors, the
detection efficiencies are recalculated using a range of
$\sigma_\mathrm{int}$ values from $0.12-0.15$.  Bins with
$\varepsilon=1$ will effectively have zero uncertainty, since the
likelihood of recovery will not depend on the details of the
population distribution; by contrast, ``low-efficiency'' bins are more
seriously affected.  These detection efficiency ``errors'' are
included into the overall rate uncertainties in \S\ref{sec:errsims}.

\subsection{Sampling time}

To remain consistent in the selection criteria used for both the
observed SNLS sample and the fake objects, we also apply the same
observational cuts described in \S\ref{sec:culls} to the artificial
SNe~Ia.  Using the peak date of each simulated light curve, we
determine whether the minimum observing requirements are met in each
filter by comparing with the SNLS image logs.  This directly
incorporates the observational cuts into the detection efficiency
calculations, while factoring in losses due to adverse weather and the
gaps between epochs.  The recovery fractions that include these
observational requirements are shown by the lower dashed lines in
Fig.~\ref{fig:compl}.

Each candidate's detection efficiency is multiplied by a factor to
account for its corresponding sampling time window for detection,
yielding a ``time corrected'' rest-frame efficiency $\varepsilon_T$:
\begin{equation}
\varepsilon_T = \varepsilon\,\frac{1}{(1+z)}\frac{\Delta T}{\mathrm{yr}}
\end{equation}
The sampling period $\Delta T$ (in years) for a given field-season is
\begin{equation}
\Delta T = \frac{1}{365.24} \left[\mathrm{max(MJD) - min(MJD)} + 30
  \right],
\label{eq:searchtime}
\end{equation}
where MJD is the modified Julian date of the available detection
images.  The extra 30 days account for the range in peak dates allowed
for the artificial SN~Ia light curves, from 20 days prior to the first
observation in a given field-season to 10 days past the final epoch.

The resulting time-corrected rest-frame detection efficiencies for the
phot-Ia sample are plotted as a function of redshift in
Fig.~\ref{fig:objeffs}. Since each field is observable for at most
$4-6$ months of the year, $\varepsilon_T$ peaks at $\sim 0.4$ even for
bright, nearby objects.

\begin{figure}
\plotone{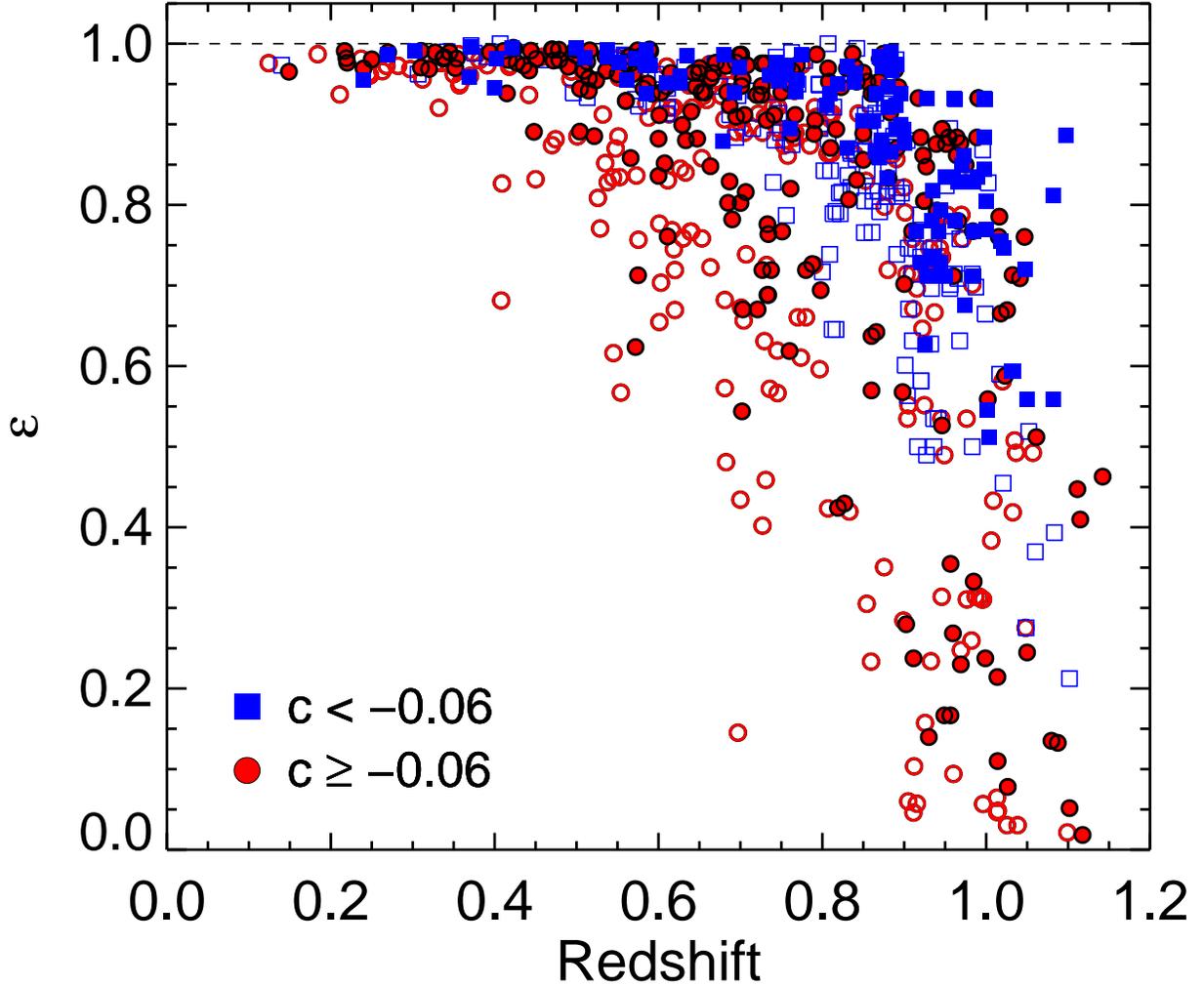}
\caption{Detection efficiencies ($\varepsilon_i$ in
  eqn.~\ref{eq:rate}) measured for each candidate in the photometric
  SN~Ia sample and plotted against redshift. SNe~Ia that are redder
  than the adopted fiducial color of $c_f \sim -0.06$ are shown as red
  circles, while bluer objects are shown as blue squares.  Open
  symbols represent SNe~Ia with stretches smaller than the median
  value of the sample ($s<1$).  These efficiencies have not been
  corrected for changes in the sampling time between the different
  fields observed.}
\label{fig:objeffs_raw}
\end{figure}

\begin{figure}
\plotone{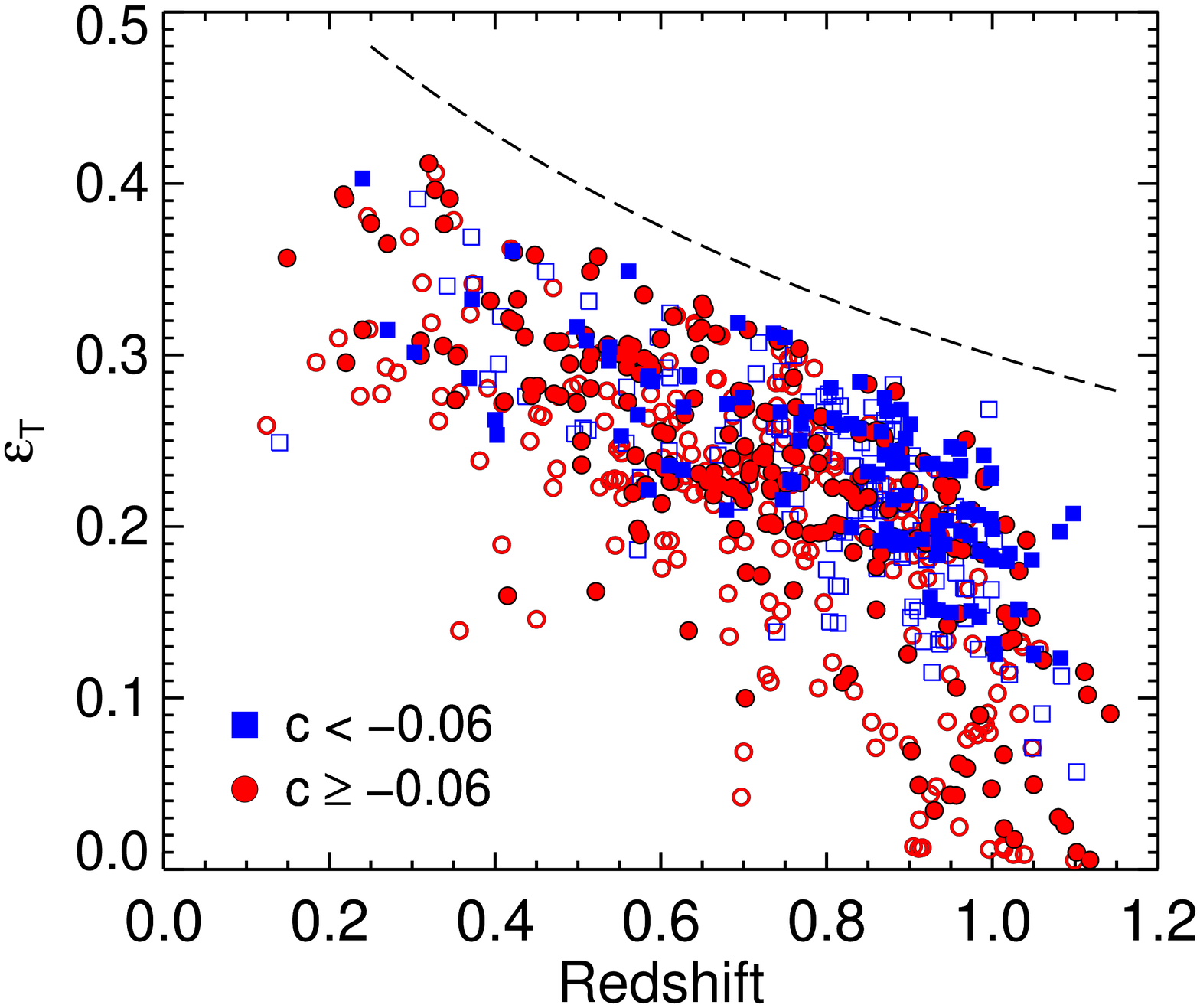}
\caption{Time-corrected rest-frame efficiencies for the SNLS phot-Ia
  sample plotted against redshift. The efficiencies shown here are
  $<1$ even at low-$z$ since they have been adjusted for field
  observability.  The dashed line shows that a $(1+z)^{-1}$ slope
  matches the general trend of the data out to $z\sim 1$, where the
  detection efficiencies begin to drop off more quickly.  SNe~Ia with
  $c>c_f$ are shown as red circles and bluer ones as blue squares.
  Lower-stretch ($s<1$) events are displayed as open symbols.  There
  are no significant differences in the median values of
  $\varepsilon_T$ as a function of redshift for high- and low-stretch
  objects.}
\label{fig:objeffs}
\end{figure}

Figs.~\ref{fig:objeffs_raw} and \ref{fig:objeffs} show that there is a
drop-off in the efficiencies above $z=0.9$, in particular for the
redder $c$ bins, making it more difficult to calculate accurate rates
at these redshifts due to color--stretch bins which are not sampled.
At $z>1.1$, it is not possible to measure SN~Ia rates using this
method due to poor survey sensitivity and inadequate statistical
sampling of spectral templates.  Therefore, we restrict our volumetric
rate calculations to the range $0.1\leq z < 1.1$.

\section{SN I\lowercase{a} rates}
\label{sec:SNLSrates}

Volumetric SN~Ia rates are calculated from eqn.~\ref{eq:rate} by
summing the observed SNe~Ia weighted by the inverse of their
time-corrected rest-frame efficiencies. The total sampling volumes for
the deep fields in redshift bins of $\Delta z=0.1$
(eqn.~\ref{eq:volume}) are provided in Column 2 of
Table~\ref{tab:rates}.  Columns 3 and 4 show the numbers of observed
candidates in each bin for the entire sample ($N_\mathrm{obs}$) and
for the spectroscopically confirmed SNe~Ia ($N_\mathrm{spec-Ia}$) in
each redshift bin.  The ``raw'' measured rates ($r_\mathrm{meas}$)
with their weighted statistical errors are given in Column 5, in units
of $\runit$; see later sections for the meaning of the remaining
columns.

\begin{deluxetable}{lccccccc}
\tabletypesize{\footnotesize}
\tablewidth{0pt}
\tablecaption{Volumetric rates from the SNLS sample\label{tab:rates}}
\tablehead{
  \colhead{$z$ bin} & 
  \colhead{Survey Volume $V$} & 
  \colhead{$N_\mathrm{obs}$} &
  \colhead{$N_\mathrm{spec-Ia}$} &
  \colhead{$r_\mathrm{meas}$} &
  \colhead{$r^{\prime}_\mathrm{meas}$} &
  \colhead{$\langle z \rangle$} & 
  \colhead{$r_\mathrm{V}$\tablenotemark{a}} \\
  \colhead{} & 
  \colhead{\footnotesize($10^4$ Mpc$^3$)} & 
  \colhead{} & 
  \colhead{} & 
  \multicolumn{2}{c}{\footnotesize($\runit$)} & 
  \colhead{} & 
  \colhead{\footnotesize($\runit$)}
}
\startdata
$0.10-0.20$ &  17.3 &   4 &  3 & $0.21 \pm 0.11$ &\nodata& 0.16 & $0.14 ^{+0.09}_{-0.09}$$^{+0.06}_{-0.12}$ \\
$0.20-0.30$ &  42.8 &  16 & 16 & $0.30 \pm 0.08$ &\nodata& 0.26 & $0.28 ^{+0.07}_{-0.07}$$^{+0.06}_{-0.07}$ \\
$0.30-0.40$ &  75.7 &  31 & 24 & $0.35 \pm 0.07$ &\nodata& 0.35 & $0.36 ^{+0.06}_{-0.06}$$^{+0.05}_{-0.06}$ \\
$0.40-0.50$ & 112.7 &  42 & 29 & $0.36 \pm 0.06$ &\nodata& 0.45 & $0.36 ^{+0.06}_{-0.06}$$^{+0.04}_{-0.05}$ \\
$0.50-0.60$ & 151.5 &  72 & 47 & $0.48 \pm 0.06$ &\nodata& 0.55 & $0.48 ^{+0.06}_{-0.06}$$^{+0.04}_{-0.05}$ \\
$0.60-0.70$ & 190.1 &  91 & 36 & $0.55 \pm 0.06$ & $0.57\pm0.06$ & 0.65 & $0.48 ^{+0.05}_{-0.05}$$^{+0.04}_{-0.06}$ \\
$0.70-0.80$ & 227.2 & 110 & 56 & $0.59 \pm 0.06$ & $0.57\pm0.06$ & 0.75 & $0.58 ^{+0.06}_{-0.06}$$^{+0.05}_{-0.07}$ \\
$0.80-0.90$ & 262.1 & 128 & 44 & $0.64 \pm 0.06$ & $0.65\pm0.06$ & 0.85 & $0.57 ^{+0.05}_{-0.05}$$^{+0.06}_{-0.07}$ \\
$0.90-1.00$ & 294.1 & 141 & 25 & $1.20 \pm 0.17$ & $0.99\pm0.29$ & 0.95 & $0.77 ^{+0.08}_{-0.08}$$^{+0.10}_{-0.12}$ \\
$1.00-1.10$\tablenotemark{b}   & 323.0 &  50 &  6 & $0.93 \pm 0.25$ & $0.51\pm0.26$ & 1.05 & $0.74 ^{+0.12}_{-0.12}$$^{+0.10}_{-0.13}$ \\
\enddata 
\tablenotetext{a}{The first error listed is statistical, and the
  second systematic.}
\tablenotetext{b}{Bins at $z>1.0$ are not
  included in the rates analysis; see Section~\ref{sec:distcorr}}
\end{deluxetable}

Contamination by non-Ias that survive the culling criteria is
estimated to contribute under $2\%$ to the total measured rates to
$z\sim 1$.  The contribution is found to be negligible up to $z\sim
0.5$, at which point it increases to around $4\%$ at $z\sim 1$.  This
is determined by summing $1/\varepsilon_T$ for the known non-Ias, and
dividing by the corresponding value for objects in each redshift bin
with available spectroscopy.  The $\varepsilon_T$ values used here are
based on the results obtained when allowing redshift to vary in the
fits, not when holding $z$ fixed at the spectroscopic redshifts.

We now correct the raw measured rates for potential systematic offsets
in the photometric redshifts and other parameters.  This is done using
the technique described next in \S\ref{sec:errsims}, which also
computes a combined statistical and systematic uncertainty on the
final rates.  The SN~Ia rates are potentially also sensitive to the
inclusion of very low detection efficiency candidates at $z\ga 0.9$,
and we must consider the effects of undetected SNe~Ia in $z/s/c$ bins
with very poor detection recovery rates.  These low-efficiency issues
are discussed later in \S\ref{sec:distcorr}.

\subsection{Error analysis}
\label{sec:errsims}

In addition to the simple ``root-$N$'' statistical errors, a number of
additional uncertainties also affect our measured rates. These can
include errors in the measured SN (photometric) redshift, stretch, and
color, which together determine the detection efficiency (and hence
weight) assigned to each event.

\begin{figure}
\plottwo{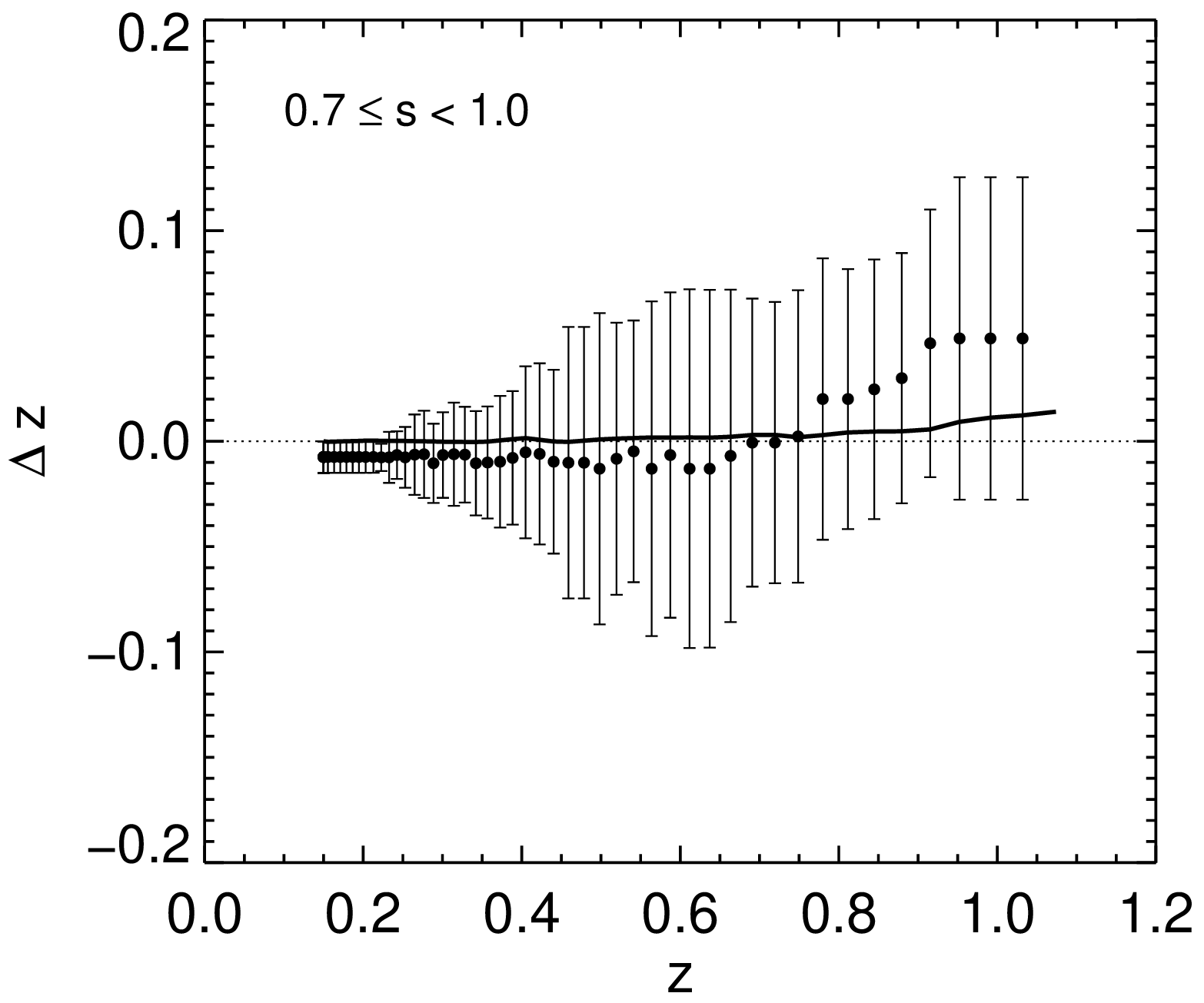}{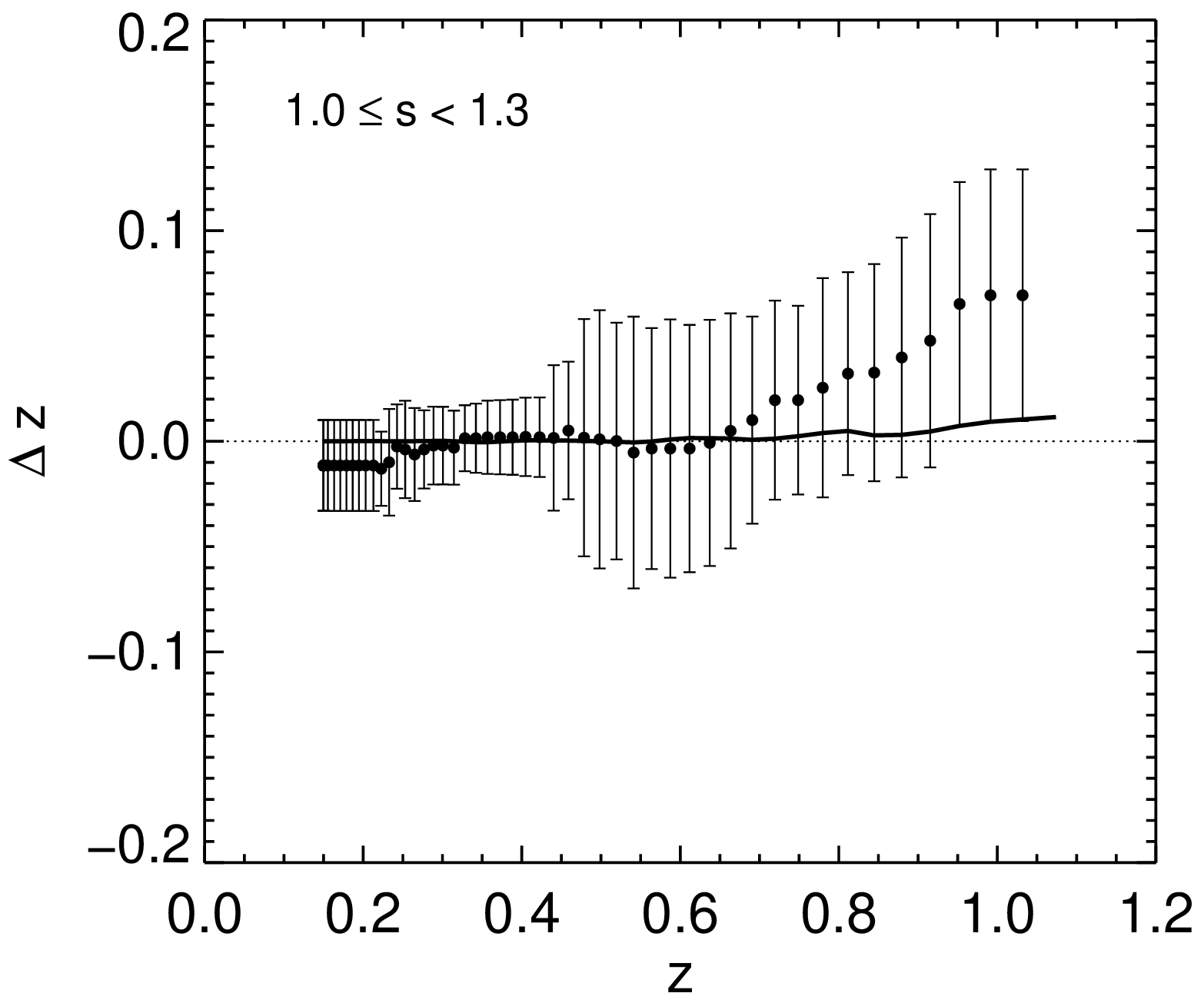}
\caption{The redshift offset, $\Delta z =
  z_{\mathrm{spec}}-z_{\mathrm{SNphot}}$, as a function of SN or host
  spectroscopic redshift for the phot-Ia sample.  The offsets are
  calculated separately in two stretch bins: $0.7\leq s < 1.0$ (upper
  panel) and $1.0\leq s < 1.3$ (lower panel).  The median $z$ offset
  in sliding bins of width $\Delta z=0.2$ are shown by the solid
  points, with error bars representing the standard deviation in each
  bin.  The offsets increase from approximately zero at $z=0.7$ to
  $\Delta z\sim 0.05-0.07$ at $z>1$.  The solid lines represent the
  expected offset due purely to sample selection bias in each stretch
  range.}
\label{fig:zerrs}
\end{figure}

As shown in Fig.~\ref{fig:zcomp}, there is a redshift uncertainty in
each measure of $z_{\mathrm{SNphot}}$.  The discrepancy between
$z_{\mathrm{spec}}$ and $z_{\mathrm{SNphot}}$ for the confirmed SN Ia
sample is presented in Fig.~\ref{fig:zerrs} for two bins in stretch.
There is a small offset above $z=0.7$, increasing to $\Delta z =
z_{\mathrm{spec}}-z_{\mathrm{SNphot}}\approx 0.05$ ($\sigma=0.08$) at
$z>1$ in the $0.7\leq s < 1.0$ sample, and $\approx 0.07$
($\sigma=0.06$) in the $1.0\leq s < 1.3$ sample.  On average, the
$z_{\mathrm{SNphot}}$ measurements are underestimated, with an
increasing offset to higher redshift.

An offset is expected based on a Malmquist bias, such that brighter
objects are more likely to have a spectroscopic type at a fixed
redshift \citep{per10}. However, we estimate this effect to be
smaller: the solid lines in Fig.~\ref{fig:zerrs} show this predicted
offset as a function of redshift.  This is calculated by estimating
the rest-frame $B$-band apparent magnitude with $z$ for the adopted
cosmology, and applying the $\Delta$mag offsets contributed by
spectroscopic selection as measured in \citet{per10}.

\begin{figure}
\plotone{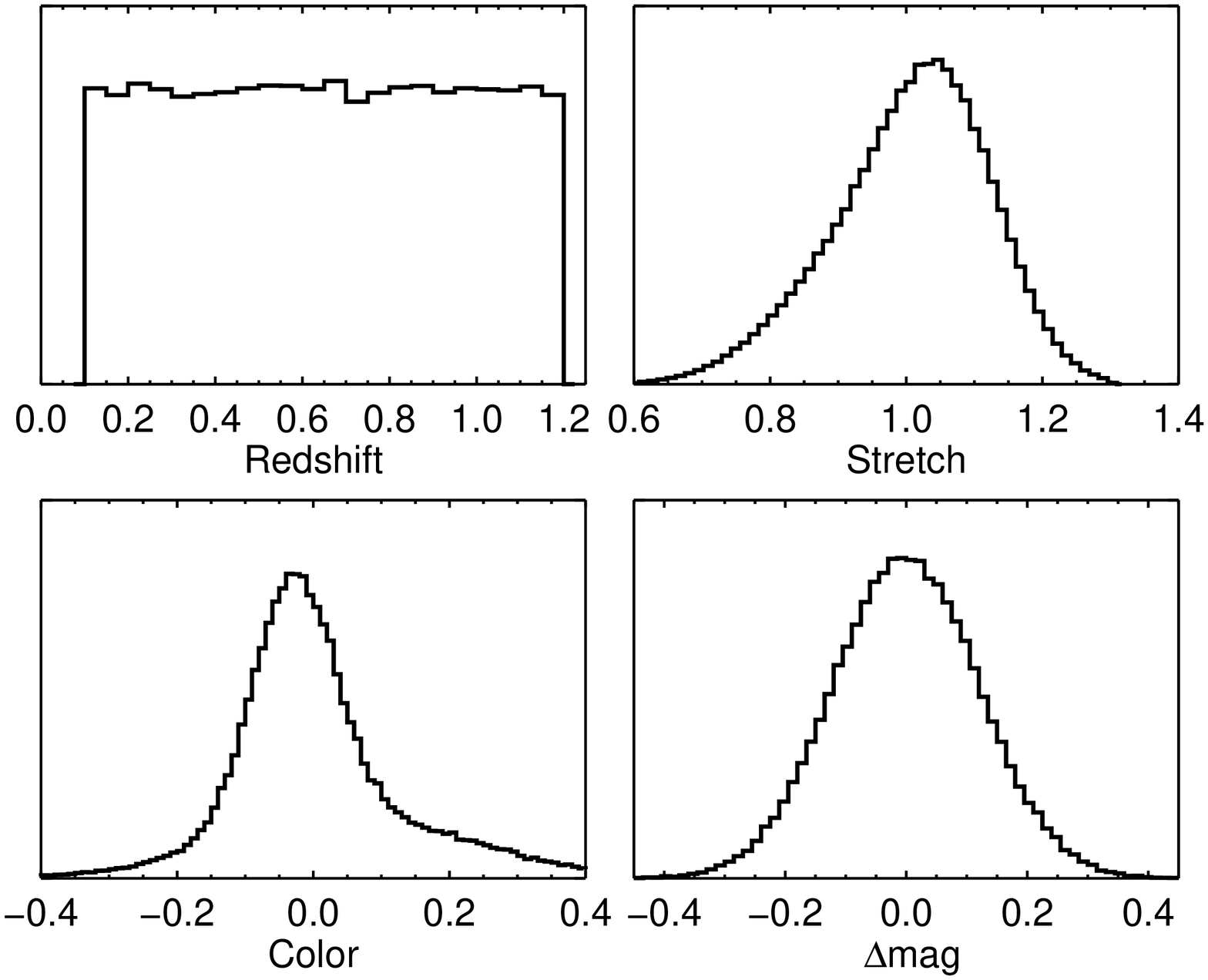}
\caption{Resampled distributions showing the properties of the
  artificial SNe~Ia used as input to the rate error simulations.
  $\Delta$mag refers to the scatter in SN~Ia rest-frame $B$-band peak
  magnitudes, and has a dispersion of $\sigma_\mathrm{int}=0.14$.}
\label{fig:simdist}
\end{figure}

To study these various uncertainties, and to handle this redshift
migration effectively, we perform a set of Monte Carlo simulations on
the measured rates.  We begin with the basic rate measurements,
$r_{\mathrm{meas}}(z)$, from Table~\ref{tab:rates}, and calculate how
many measured SNe~Ia that rate represents in each redshift bin by
multiplying by the volume in that bin: $N_{\mathrm{meas}}(z)$.  Many
realizations (5000) are performed by drawing $N_{\mathrm{meas}}$
objects from typical SNLS-like distributions of artificial SNe~Ia.
These are the same artificial objects as used in the detection
efficiency calculations (\S\ref{sec:effs}), although the stretch,
color, and $\Delta$mag distributions are matched to those of the
spectroscopically confirmed SN~Ia sample \citep[see][]{per10}.  The
distributions for input to the error calculations are shown in
Fig.~\ref{fig:simdist}. Only a fraction of the objects in each
redshift bin have a spectroscopic redshift, with the remainder having
a SN Ia photometric redshift and accompanying uncertainty
(Fig.~\ref{fig:zerrs}).  This ``spectroscopic fraction'',
$F_{\mathrm{spec}}(z)$ is calculated in each redshift bin from
Fig.~\ref{fig:objpars}.

\begin{figure*}
\includegraphics[width=0.49\textwidth]{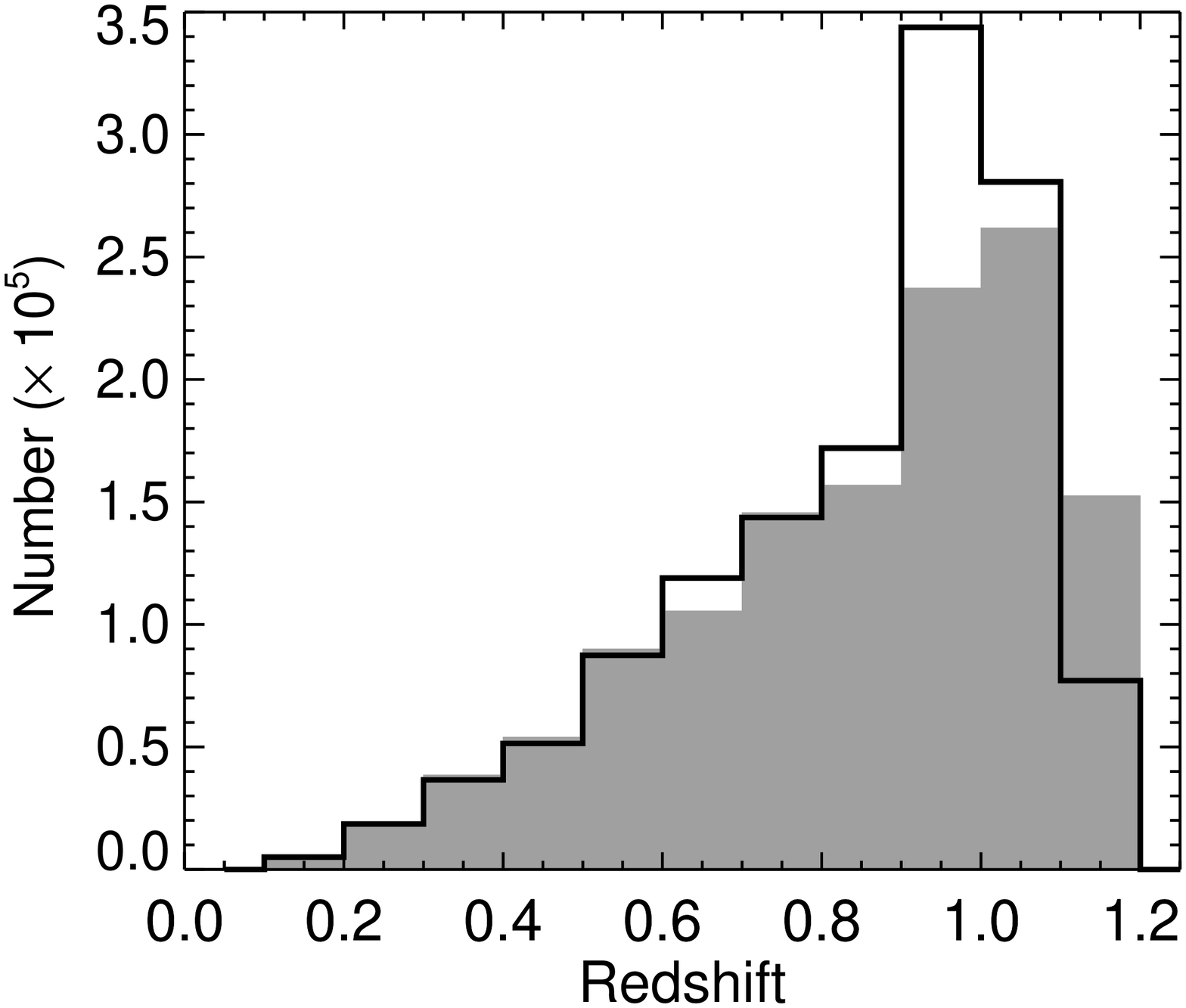}
\includegraphics[width=0.49\textwidth]{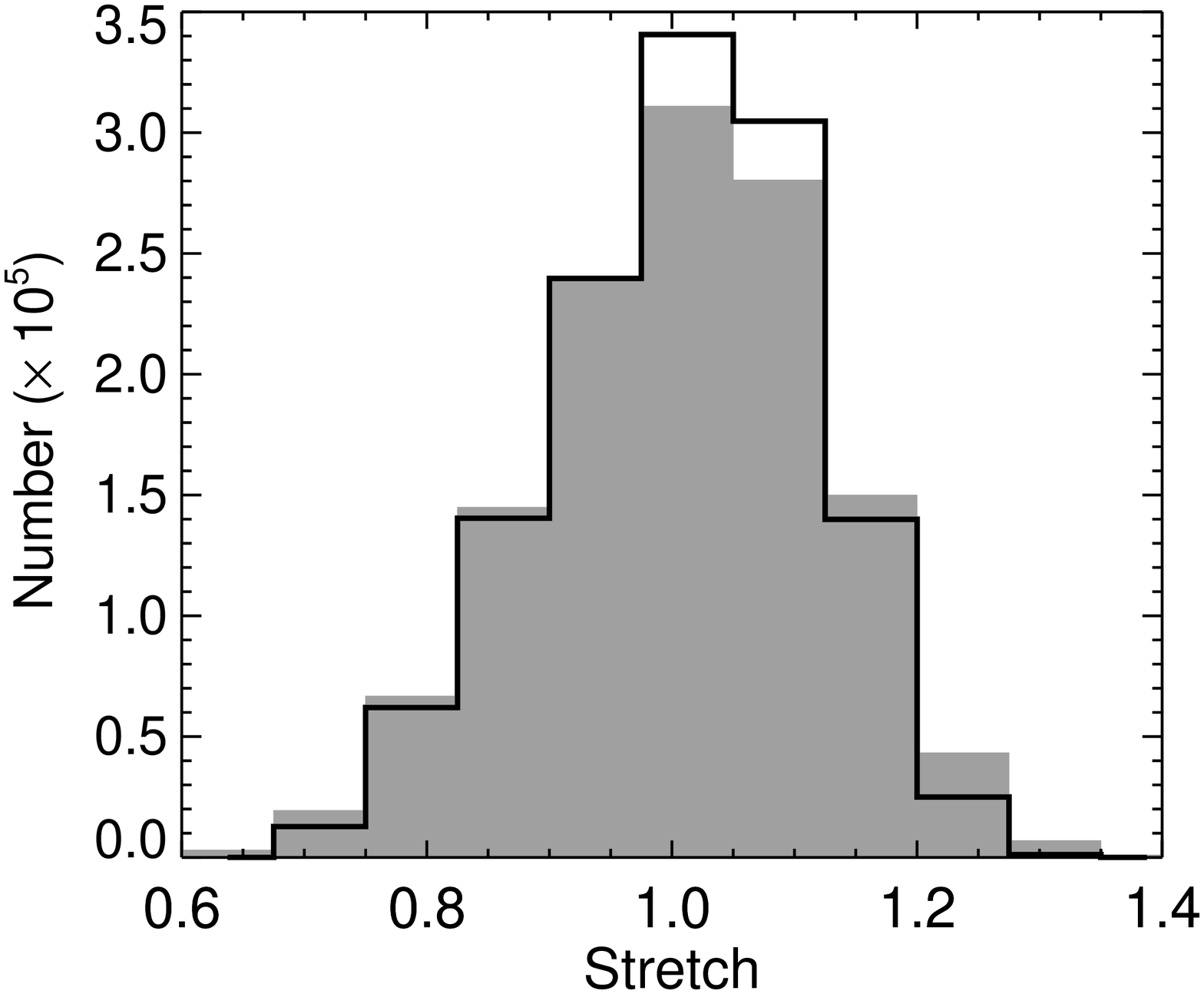}
\includegraphics[width=0.49\textwidth]{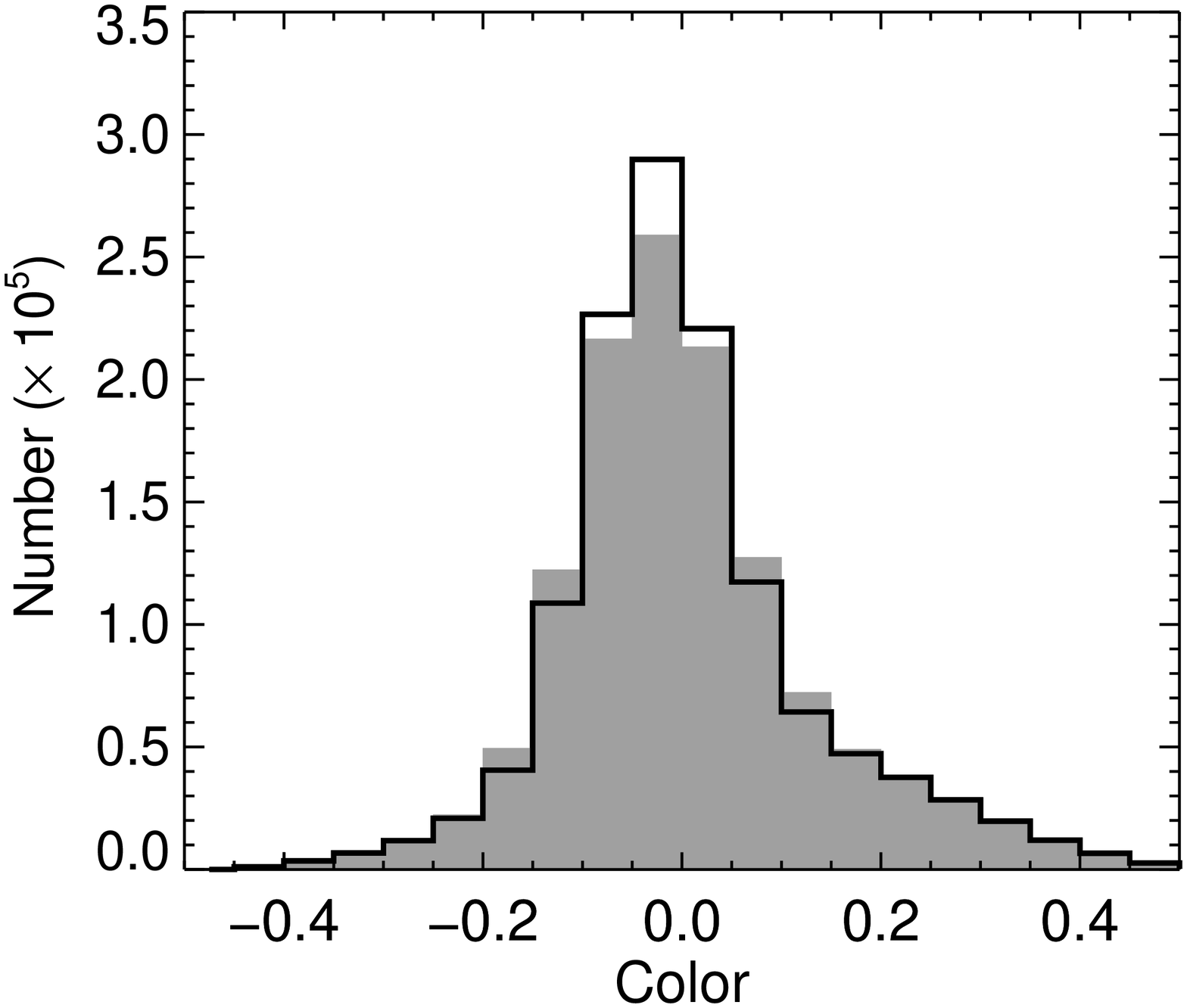}
\caption{Histograms showing the input (black line) and output (gray
  filled) parameter distributions from the rate error simulations as a
  function of redshift (left), stretch (center), and color (right).}
\label{fig:parshift}
\end{figure*}

The procedure for each realization is then as follows:

\begin{enumerate}
\item $N_{\mathrm{meas}}(z)$ is randomized according to the Poisson
  distribution, using the Poisson error based on $N_{\mathrm{obs}}(z)$
  but scaled to $N_{\mathrm{meas}}(z)$, to give $N_{\mathrm{rand}}(z)$
  simulated objects.
\item Each simulated object $i$ in each redshift bin is assigned a
  random redshift appropriate for that bin ($z_i$). Within each bin,
  the probability follows a scaled number-density profile according to
  the expected increase in volume with redshift.
\item The $z_i$ are then randomly matched to a SN Ia from the
  artificial distribution with the same redshift
  (Fig.~\ref{fig:simdist}), and that event's stretch ($s_i$) and color
  ($c_i$) assigned to the simulated event.
\item Using the fraction of spectroscopic redshifts in each bin
  $F_{\mathrm{spec}}(z)$ (Fig.~\ref{fig:objpars}), we assign this
  fraction of the $N_{\mathrm{rand}}$ objects to correspond to a
  spectroscopic redshift measurement.  The remaining redshifts are
  assumed to come from a photometric fit, and are shifted and
  randomized using the median offsets and standard deviations shown in
  Fig.~\ref{fig:zerrs} to give $z^{\prime}_i$.
  Any event with a $z_{\mathrm{spec}}$ is not adjusted.
\item Correlated stretch and color errors are then incorporated for
  all objects, using typical covariances produced by SiFTO for the
  SNLS sample, and the $s_i$ and $c_i$ randomized to $s^{\prime}_i$ and
  $c^{\prime}_i$.
\item $z^{\prime}_i$, $s^{\prime}_i$, and $c^{\prime}_i$ are used to
  match each simulated object to a detection efficiency.  Efficiency
  errors are included by applying a random shift drawn from an
  two-sided Gaussian representing the asymmetric uncertainties on each
  value (\S\ref{sec:MC}).
\item Random numbers between zero and one are generated to evaluate
  whether each simulated object gets ``found'': if the selected number
  is lower than the detection efficiency associated with the simulated
  object, that event is added to the rate calculated for that
  iteration.
\item The rate for that iteration is then calculated using the
  appropriate detection efficiencies from step 6.
\end{enumerate}

The final volumetric rates are presented in Columns 7 of
Table~\ref{tab:rates}. These are calculated as the mean of the 5000
simulated rates in each redshift bin. We also calculate the standard
deviation in each bin, subtract from this in quadrature the
statistical Poisson uncertainty based on $N_{\mathrm{obs}}(z)$, with
the remainder our estimate of the systematic uncertainty in each
redshift bin.

The effects of the simulations described above on the input redshift,
stretch, and color distributions are shown in Fig.~\ref{fig:parshift}.
The redshift histogram shows that the offsets applied in step~4 of the
simulations produce a net increase in redshift to compensate for the
small bias in the photometric fitting, with the effect increasing
towards higher $z$.  This causes a flattening of the output rates
calculated by the simulations as compared with the measured (and
uncorrected) rates (Table~\ref{tab:rates}).  In addition to the offset
towards higher redshifts at $z>0.7$, there is also a very small
spread in the stretch and color distributions
(Fig.~\ref{fig:parshift}).


\subsection{Low-efficiency candidates}
\label{sec:distcorr}

The detection efficiencies in some of the reddest $c$ bins begin to
rapidly decrease at $z\ga 0.9$ (Fig.~\ref{fig:objeffs}).  As the
contribution to the rate from each observed SN~Ia goes as
$1/\varepsilon_T$, the measured volumetric rates are particularly
sensitive to any objects with very low detection efficiencies
($\varepsilon_T$). There is also the potential for a complete omission
of SNe Ia in some $z/s/c$ bins.  For example, in $z/s/c$ bins with
less than $10\%$ detection efficiency, on average at least 10 SNe~Ia
must be observable in a given field-season for just one to be
detected.  If that one SN is not detected by the real-time pipeline,
the 10 SNe~Ia that it truly represents in that bin will never be
counted in the final rates tally (and the rate measurement will be
biased).

To examine the sensitivity of our rates on the ``low-$\varepsilon_T$
regions'' of $z/s/c$ parameter space, we use the $z<0.6$
detection-efficiency-corrected SNLS sample as a model for the true
($s$, $c$) SN~Ia distribution at $z>0.6$.  This population is assumed
observationally complete (Fig.~\ref{fig:compl}) and is taken to be
representative of the underlying sample of SNe~Ia in the
universe\footnote{Of course, this relies on the (possibly incorrect)
  assumption of no evolution in intrinsic stretch or color as a
  function of redshift \citep[e.g.,][]{how07}.}.  The two-dimensional
($s$, $c$) distribution at $z<0.6$ is fit to the five $z>0.6$ bins,
and the best-fit scaling determined.  The total rates
$r^{\prime}_{\mathrm{meas}}(z)$ are then calculated from those scaled
numbers (tabulated in Table~\ref{tab:rates})

These tests indicate that, while the results remain consistent within
their errors up to $z=0.95$ ($0.99 \pm 0.29 \runit$ compared with
$r_\mathrm{V}=1.20 \pm 0.17 \runit$), there is a significant amount of
uncertainty in the SNLS rates at higher redshifts due to sample
incompleteness.  At $z=1.05$, the scaled rate is $0.51 \pm 0.26
\runit$, whereas our the calculated value is $r_\mathrm{V}= 0.93 \pm
0.25 \runit$.  This finding is consistent with the results shown in
Figs.~\ref{fig:objpars} and \ref{fig:objeffs_raw}: the phot-Ia sample
numbers drop significantly at $z>1.0$, and those that are found in the
sample can have very low $\varepsilon_T$.  For these reasons, we limit
the formal analysis of the SNLS rates to $r_\mathrm{V}(z < 1.0)$.

\section{Delay-time distributions}
\label{sec:dtds}

Having measured volumetric SN Ia rates and associated errors, we now
compare our measurements with those of other studies.  We also examine
the SN Ia rate evolution as a function of redshift, and compare with
predictions based on various simple delay-time distribution models
from the literature.

For comparison of the SNLS rate measurements to various SN Ia models,
additional data in redshift ranges not sampled by SNLS is required. In
the rest of this section, we will make use of an extended SN Ia rate
sample comprising the \citet{li11b} LOSS measurement at $z\sim0$, the
\citet{dil10} sample from SDSS-SN at $z\sim0.2$, and the recent
\citet{gra11} Subaru Deep Field (SDF) sample at higher redshifts,
together with our SNLS results.  Clearly other samples could have been
chosen -- however, these three are the largest SN Ia samples in their
respective redshift ranges, and have the greatest statistical power.
In the case of the SDSS and SDF samples, they are also built on
rolling SN searches similar to SNLS.

We make some small corrections to these published rates in order to
ensure a fair comparison across samples. The \citet{li11b} sample
includes all sub-classes of SNe Ia, including the peculiar events in
the SN2002cx-like class and sub-luminous events in the SN1991bg-like
class. SN2002cx-like events make up 5\% of the LOSS volume-limited SN
Ia sample \citep{li11a}. These are not present (or accounted for) in
the SNLS sample, and are excluded from the SDSS analysis
\citep{dil08}, so we therefore exclude these from the \citet{li11b}
sample, reducing their published rate value by 5\%.

Both the \citet{li11b} and \citet{dil10} samples include SNe Ia in the
sub-luminous SN1991bg category \citep[see also][]{dil08}, which we
exclude here in the SNLS analysis \citep[these are studied
in][]{gon11}. While we could correct our own rates for this population
using the \citet{gon11} results, it is unclear how to treat the SDF
sample in the same way (do SN1991bg-like events even occur at $z>1$?)
and the SNLS sub-luminous measurement is quite noisy. Instead we use
the very well-measured fraction of SN1991bg-like SNe in the
volume-limited LOSS sample (15\%), and reduce both the LOSS and SDSS
published rates by this amount. This 15\% is based on the
classifications given in \citet{li11b}. We confirm that this is
appropriate for our stretch selection (i.e., we require $s>0.7$) by
fitting the available \citet{li11b} light curves with SiFTO. 16\% of
the available \citet{li11b} sample has a fitted stretch $<0.7$,
consistent with the 15\% reported as 91bg-like by \citet{li11b}.

\subsection{Comparison with published rates}

\begin{figure}
\plotone{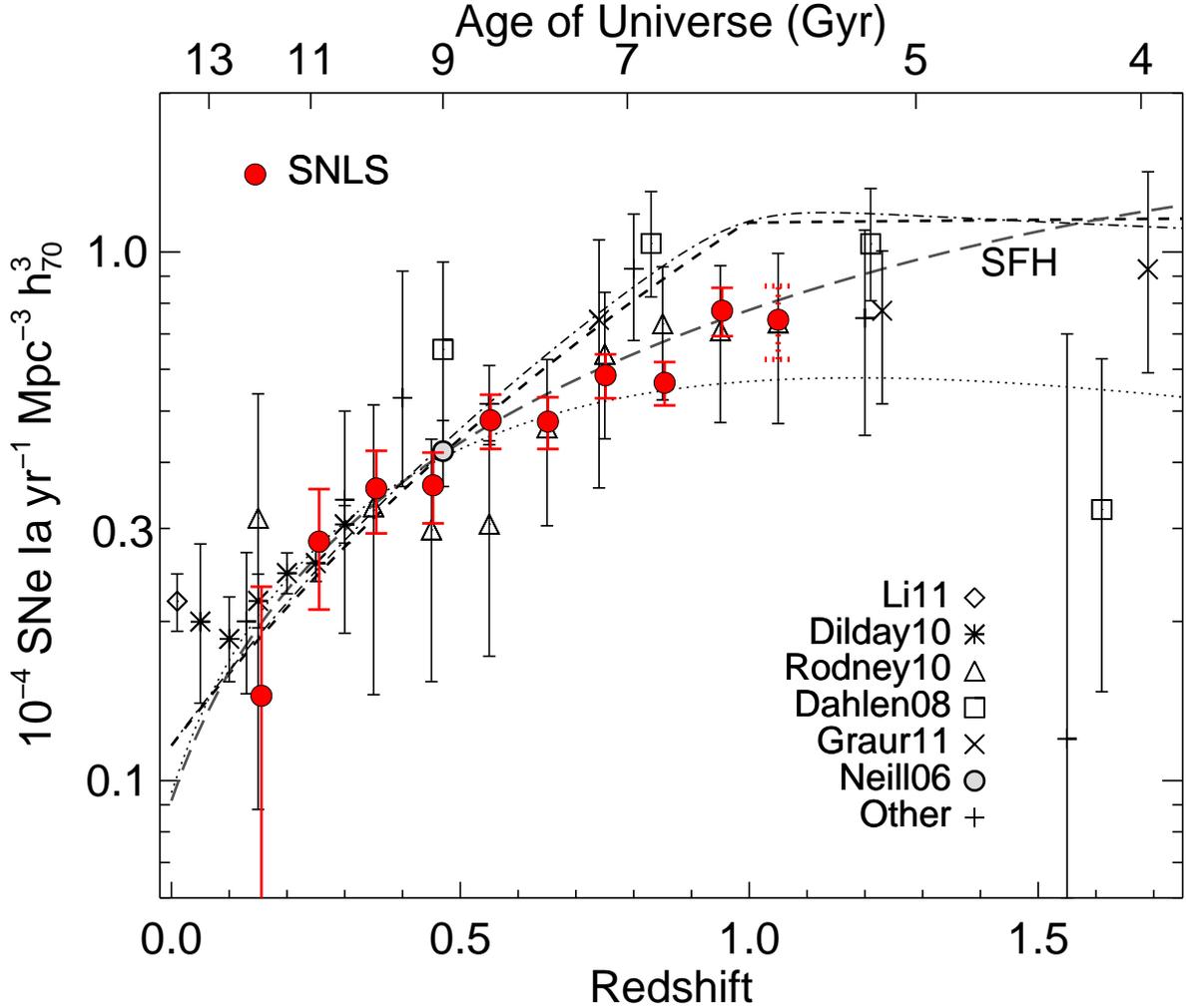}
\caption{The SNLS volumetric SN~Ia rates in the context of the data in
  Fig.~\ref{fig:rates_lit}.  The filled circles represent the SNLS
  rates from the current analysis.  The rate at $z=1.05$ (with the
  dashed error bar) represents the redshift bin in which
  incompleteness and poor spectroscopic sampling make measurements
  untrustworthy.  SNLS rates above $z=1.0$ are not included in
  subsequent fits. The samples of \citet{li11b} and \citet{dil10} have
  been scaled downwards to reflect the exclusion of sub-luminous and
  SN2002cx-like SNe Ia from the SNLS sample (see $\S$~\ref{sec:dtds}).
  Over-plotted are the various SFHs we use in our analysis in
  $\S$~\ref{sec:dtds} as fit to the SNLS data only. The short-dashed
  line shows the piece-wise SFH from \citet{li08}, the long-dashed line
  the Cole et al. form from \citet{li08}, the dot-dashed line the SFH
  from \citet{yuk08}, and the dotted line the SFH of \citet{wil08}.}
\label{fig:rates_comp}
\end{figure}

Fig.~\ref{fig:rates_comp} shows the SNLS volumetric SN~Ia rates for
comparison with recent published results. The SNLS volumetric rate at
$z\sim 0.5$ published by \citet{nei06} is $\snr(\langle
z\rangle=0.47)=[0.42^{+0.13}_{-0.09}\mathrm{(syst)}\pm
0.06\mathrm{(stat)}]$\runit, consistent with our binned rates in the
same redshift range.  Note that the effects of non-Ia contamination
(\S\ref{sec:SNLSrates}) have not been incorporated into the SNLS
errors shown in Fig.~\ref{fig:rates_comp}. Our results are also
consistent with \cite{dil10} at $<0.3$, and with \citet{rod10} at
higher redshifts, although those latter measurements have
significantly greater uncertainties.

\begin{figure}
\plotone{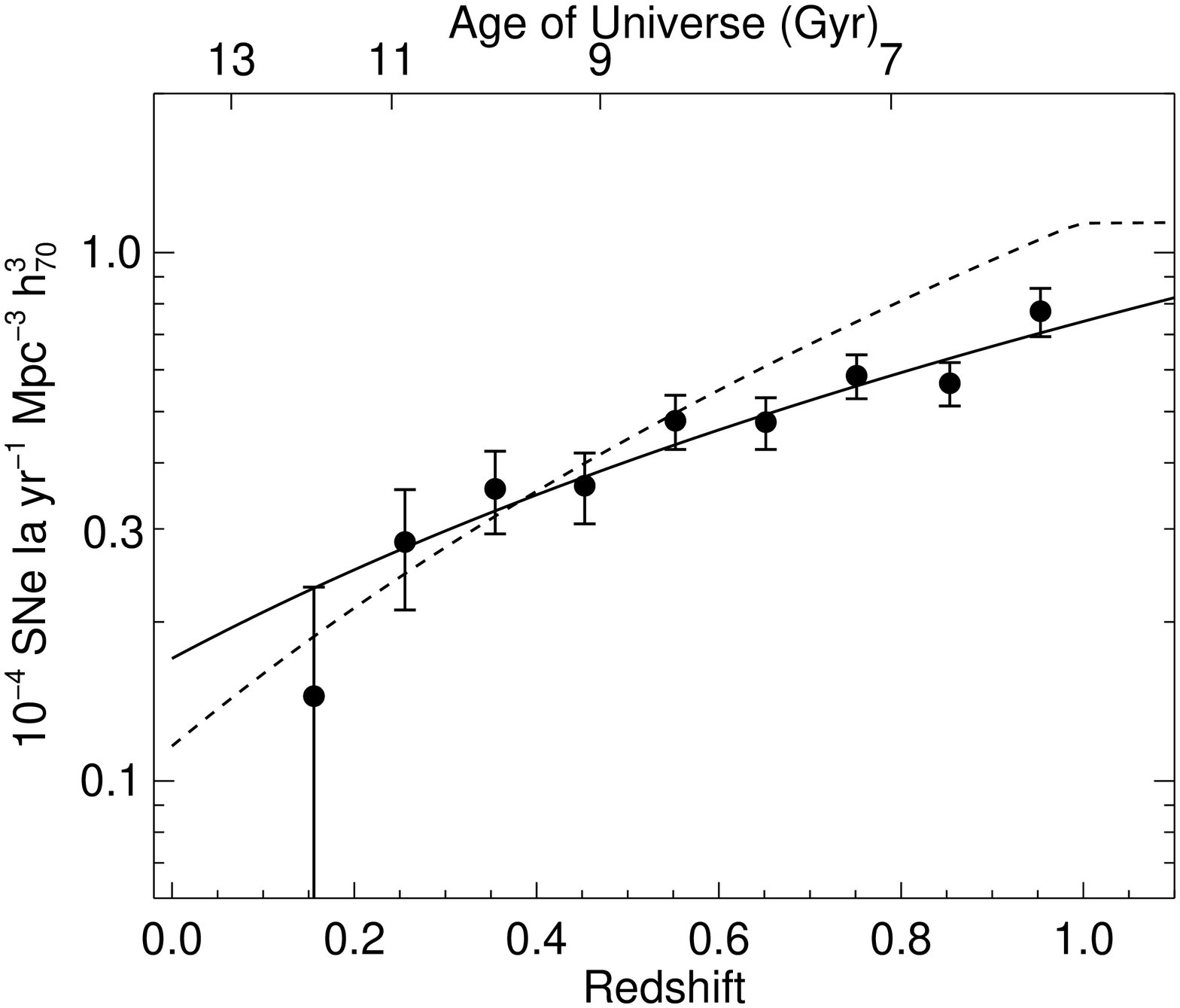}
\caption{SNLS rates as a function of redshift, showing a power-law fit
  to the data (solid line): $\snr(z)=r_0(1+z)^\alpha$, where
  $\alpha=2.11\pm 0.28$ and $r_0=(0.17 \pm 0.03)\runit$.  The reduced
  $\chi^2$ goodness-of-fit statistic is $\chi^2_{\nu}=0.64$.  For
  comparison, the dashed line shows the \citet{col01} form of the
  \citet{li08} star formation history profile, which has $\alpha=3.3$
  out to $z\sim1$.}
\label{fig:rates_powerlaw}
\end{figure}

The SNLS rates show a rise out to $z\sim 1$, with no evidence of a
rollover at $z\sim 0.5$.  We can parameterize the SNLS rate evolution
as a simple power-law:
\begin{equation}
\snr(z)=r_0(1+z)^\alpha,
\end{equation}
with the best-fit shown as the solid line in
Fig.~\ref{fig:rates_powerlaw}. We find $\alpha=2.11\pm 0.28$ and
$r_0=(0.17 \pm 0.03)\runit$ (all errors in this section are
statistical only). This evolution is shallower than a typical fit to
the cosmic star-formation history, with $\alpha\simeq3.3$
\citep[e.g.][]{li08}, though the constraining power of the SNLS data
alone at $z<0.3$ is not great. By comparison, \citet{dil10} find
$r_0=(0.23\pm0.01)\runit$ and $\alpha=2.04^{+0.90}_{-0.89}$ using the
lower-redshift SDSS-SN data, completely consistent with our results.
Including all the external data gives $r_0=(0.21 \pm 0.01)\runit$ and
$\alpha=1.70\pm0.12$.

\subsection{Comparison with delay-time distribution models}
\label{sec:comp-with-models}

\begin{deluxetable}{lccccccccccc}
\tabletypesize{\scriptsize}
\setlength{\tabcolsep}{0.025in} 
\tablecolumns{12}
\tablecaption{Various DTD model fits to the volumetric SN Ia rate, \snr.\label{tab:ABchi2}}
\tablehead{
 &&&\multicolumn{3}{c}{\citet{li08} piece-wise SFH}&\multicolumn{3}{c}{\citet{li08} ``Cole et al.'' SFH}&\multicolumn{3}{c}{\citet{yuk08} SFH}\\
  \colhead{Data} & 
  \colhead{Model} &\colhead{$\nu$} &
  \multicolumn{2}{c}{Fit parameters}&
  \colhead{$\chi^2_{\nu}$}&
  \multicolumn{2}{c}{Fit parameters}&
  \colhead{$\chi^2_{\nu}$}&
  \multicolumn{2}{c}{Fit parameters}&
  \colhead{$\chi^2_{\nu}$}}
\startdata
               &                        &     & $\tau$ (Gyr)         &    &          & $\tau$ (Gyr)                     &       &                   & $\tau$ (Gyr)                     &        &                  \\[2pt]
\tableline                                                                                                                                                
Extended\tablenotemark{a}  & Gaussian               & 18  & 3.4        & \nodata      & 2.62    & 3.4                                & \nodata      & 1.60 &  3.4                                & \nodata      & 3.29   \\
Extended  & Gaussian               & 17  & $3.1\pm0.3$        & \nodata      & 2.64    &   $2.5\pm0.7$                             & \nodata      & 1.60&   $2.9\pm0.3$                             & \nodata      & 3.00   \\
Ext.+D08  & Gaussian        & 19  & 3.4        & \nodata      & 2.72    & 3.4                                & \nodata      & 1.68 &   3.4                                & \nodata      & 3.43   \\
\tableline\\[-5pt]                                                                                                                                        
               &                        &     & $\beta$        &              &      & $\beta$           &           &        & $\beta$             &                          &        \\[2pt]
\tableline                                                                                                                                                
SNLS      & Power law              & 8  & 1    &\nodata             & 0.76    & 1                 & \nodata                       &1.36 &    1                                  & \nodata      &0.81    \\
Extended  & Power law              & 18  & 1    &\nodata             & 0.72    & 1                                  & \nodata      &1.06 &    1                                  & \nodata      &0.78    \\
Extended  & Power law              & 17  & $0.98\pm0.05$  &\nodata             & 0.76    & $1.15\pm0.08$                               & \nodata      & 0.92 &   $0.98\pm0.05$                & \nodata      & 0.81   \\
\tableline\\[-5pt]                                                                                                                                        
               &                        &     & $\tau$ (Gyr)        &           &         & $\tau$ (Gyr)       &                 &        & $\tau$ (Gyr)        &                  &        \\[2pt]
\tableline                                                                                                                                                
SNLS & Exponential & 7  & $1.5\pm0.4$  &\nodata        & 0.85    & $0.2\pm2.9$                               &\nodata & 0.58 &   $1.4\pm0.4$                               &\nodata & 0.91   \\
Extended & Exponential & 17  & $2.6\pm0.3$  &\nodata        & 1.33    & $2.1\pm0.3$                               &\nodata & 1.15 &   $2.5\pm0.3$                               &\nodata & 1.44   \\
\tableline\\[-5pt]                                                                                                                                                
               &                        &     & $\beta$     &              &         & $\beta$                  &                          &        & $\beta$                  &                          &        \\[2pt]
\tableline                                                                                                                                                
SNLS      & \citetalias{pri08}     & 8  & 0.5                     &\nodata & 4.14        &0.5               &\nodata                    & 4.81 &  0.5                                  &\nodata & 4.31   \\
Extended  & \citetalias{pri08}     & 18  & 0.5                     &\nodata & 4.08        &0.5               &\nodata                    & 4.77  & 0.5               &\nodata                    & 4.19   \\
\tableline\\[-5pt]                                                                                                                                        
               &                        &     & $A$\tablenotemark{a}            & $B$\tablenotemark{b}        &         & $A$   & $B$          &     & $A$   & $B$ &    \\[2pt]
\tableline                                                                                                                                                
SNLS      & $A+B$                    & 7  & $1.6\pm0.5$ & $3.4\pm0.3$  & 0.74    & $0.3\pm0.6$               & $5.2\pm0.5$             &  0.59 &  $1.7\pm0.5$               & $3.2\pm0.3$             &  0.78  \\
Extended  & $A+B$                    & 17  & $1.9\pm0.1$ & $3.3\pm0.2$  & 0.60    & $1.5\pm0.2$               & $4.3\pm0.3$             &  0.77 &  $2.0\pm0.1$               & $3.1\pm0.2$             &  0.63  \\
SNLS      & $A=0$            & 8  & 0     & $4.4\pm0.3$  & 1.81    & 0                   & $5.4\pm0.2$             &  0.54 &  0 & $4.2\pm0.3$             &  1.92  \\
Extended  & $A=0$            & 18  & 0     & $5.3\pm0.4$  & 6.22    & 0                   & $6.1\pm0.3$             &  2.95 &  0 & $5.1\pm0.4$             &  6.70  \\
SNLS      & $B=0$            & 8  & $5.6\pm0.8$ & 0  & 9.72    & $6.1\pm0.9$ & 0            &  9.74 &  $5.8\pm0.8$ & 0            &  10.09  \\
Extended  & $B=0$            & 18  & $3.8\pm0.4$ & 0  & 9.42    & $4.0\pm0.4$  & 0           &  9.56 &  $3.8\pm0.4$ & 0             &  9.66  \\
\tableline\\[-5pt]                                                                                                                                        
               &                        &     & $\Psi_1$\tablenotemark{c}            & $\Psi_2$         &         & $\Psi_1$   & $\Psi_2$        &         & $\Psi_1$   & $\Psi_2$&        \\[2pt]
\tableline                                                                                                                                                
Extended  & 2-bin\tablenotemark{d}                  & 17  & $90\pm5.2$ & $1.2\pm0.10$  & 0.64    & $120\pm7.8$               & $0.88\pm0.14$             &  0.79 &  $86\pm5.0$               & $1.3\pm0.10$             &  0.67  \\

\enddata
\tablenotetext{a}{The extended sample refers to the SNLS sample plus the external data described in $\S$~\ref{sec:dtds}.}
\tablenotetext{b}{Units of \Aunit, where \msun\ refers to the current stellar mass, \mstellar.}
\tablenotetext{c}{Units of \Bunit.}
\tablenotetext{d}{Units of \Aunit, where \msun\ refers to the total formed stellar mass, \mstar.}
\tablenotetext{e}{A discrete DTD, equal to $\Psi_1$ at $t<420$Myr, and $\Psi_2$ otherwise.}
\end{deluxetable}

We now compare our rate evolution with simple parameterizations of the
delay-time distribution (DTD) from the literature relating the cosmic
star-formation history (SFH) to SN Ia rates. The DTD, $\Psi(t)$, gives
the SN Ia rate as a function of time for a simple stellar population
(SSP), i.e., following a $\delta$-function burst of star formation.
The SN Ia rate (\snr) at time $t$ is then
\begin{equation}
\label{eq:convol}
  \snr(t)=\int^t_0\mathrm{SFR}(t-\tau)\Psi(\tau)d\tau .
\end{equation}
where SFR($t$) is the star-formation rate as a function of time. Thus,
different functional forms of the DTD can be tested against
observations of volumetric rates if the SFR($t$), or the cosmic SFH,
is known \citep[e.g.,][]{mad98,str04,oda08,hb10,gra11}. An implicit
assumption in this test is that the DTD is invariant with redshift and
environment.

As our default SFH model, we choose the \citet{li08} update to the
\citet{hb06} fit to a compilation of recent star-formation density
measures. For simplicity we use a \citet{sal55} initial mass function
(IMF) with mass cut-offs at 0.1\msun\ and 100\msun, and assume that
stars began to form at $z=10$.  The default SFH is parameterized in a
piece-wise fashion, and is over-plotted on the data in
Fig.~\ref{fig:rates_comp}. As shown by \citet{for06} and
\citet{gra11}, the choice of SFH can add an additional significant
systematic uncertainty in any comparisons of DTDs to SN Ia volumetric
data. We therefore compare with results obtained using an alternative
parameterization of the SFH by \citet{li08} following \citet{col01}, as
well as a SFH fit to slightly different data by \citet{yuk08}
\cite[see also][]{hb10}. In $\S$~\ref{sec:comparison-other-dtd}, we
also investigate a SFH derived in a very different manner
\citep{wil08}.

The integral of the DTD gives the total number of SNe Ia per formed
stellar mass, $N_{\mathrm{Ia}}/\mstar$. This can be converted into the
fraction of intermediate-mass stars that explode as SNe Ia, $\eta$, by
multiplying by a factor of 47.2 (for the Salpeter IMF). This assumes
that the progenitor mass range for a SN Ia is 3--8\msun\
\citep[see][for a discussion]{mao08}. For all our model DTDs, we set
the DTD to zero at epochs earlier than 40Myr, the approximate lifetime
of an $8\msun$ star.

\subsubsection{Gaussian DTDs}
\label{sec:gaussian-dtds}

\begin{figure*}
\plotone{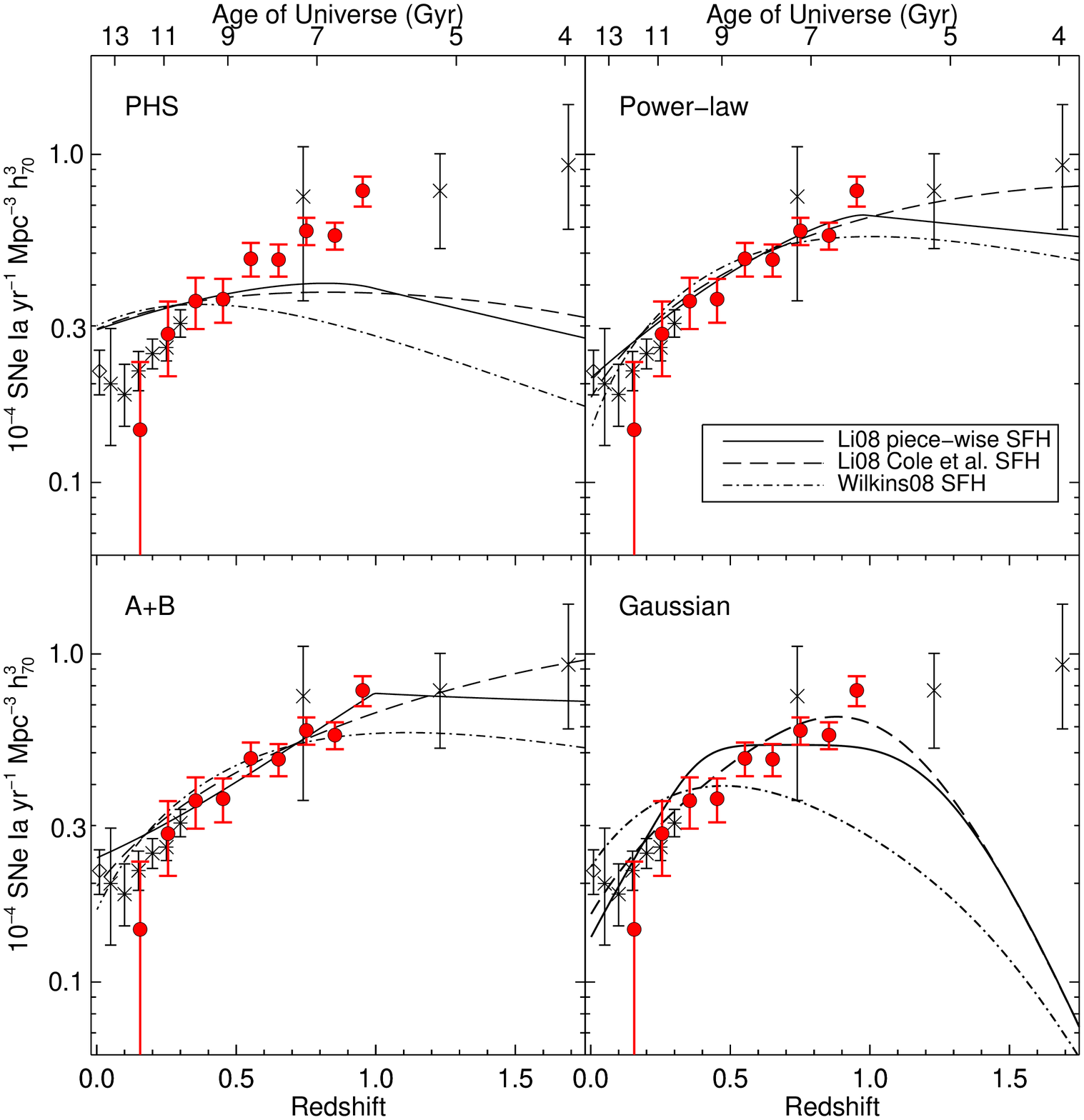}
\caption{SN~Ia rates as a function of redshift with various delay-time
  distribution (DTD) model predictions fit to the data for different
  cosmic SFHs.  Lower left: ``$A+B$'' model ($\S$~\ref{sec:aplusb}),
  lower right: Gaussian DTDs ($\S$~\ref{sec:gaussian-dtds}), upper
  left: the model of \citet{pri08} ($\S$~\ref{sec:power-law-dtds}),
  and upper right: a generic power law DTD
  ($\S$~\ref{sec:power-law-dtds}).  In all cases, solid lines
  represent the piece-wise cosmic SFH of \citet{li08}, dashed lines
  the \citet{col01} form of the \citet{li08} SFH, and dot-dash lines
  the \citet{wil08} SFH. See text for more details of the models, and
  Table~\ref{tab:ABchi2} for the numerical values.}
\label{fig:rates_dtd_compare}
\end{figure*}

We begin by fitting a Gaussian DTD, with $\Psi(t)\propto
e^{-(t-\tau)^2/(2\sigma^2)}$, to the volumetric \snr\ data, following
\citet{str04} \citep[see also][]{str10}. We fit a DTD with parameters
fixed at $\tau=3.4$\,Gyr and $\sigma=0.2\tau$ (i.e., just adjusting
the normalization in the fits), as well as a DTD fit with $\tau$
allowed to vary. The results are listed in Table~\ref{tab:ABchi2} and
compared to other DTD fits in Fig.~\ref{fig:rates_dtd_compare}.

This model has $\chi^2_{\nu}=2.62$ ($\chi^2_{\nu}$ is the reduced
$\chi^2$, the $\chi^2$ per degree of freedom, $\nu$). Allowing $\tau$
as a free parameter in the fits gives $\tau=3.1\pm0.3$\,Gyr with a
similar $\chi^2_{\nu}$ -- the fit quality is slightly better when
using Cole et al. form of the SFH ($\chi^2_{\nu}=1.60$).

The Gaussian DTDs therefore provide poor fits to the \snr\ data, and
are not capable of matching the SNLS, SDF and SDSS/LOSS data
simultaneously. In particular, these Gaussian DTDs predict a decrease
in the number of SNe at $z>1$ not seen in the combined data set.
However, they were originally favored following fits to data including
$z>1$ points from \textit{HST} searches \citep{dah04,dah08} which are
not included in our analysis due to their lower statistical precision
compared to the SDF study. As a consistency check we also replace the
SDF data with the \citet{dah08} data (adjusted downwards by 15\% to
account for 19bg-lie events) in our fits -- we find that the
$\chi^2_{\nu}$ does not improve (Table~\ref{tab:ABchi2}).

\subsubsection{Power law and exponential DTDs}
\label{sec:power-law-dtds}

Theoretically, if SNe Ia are dominated by a single channel, the DTD
will likely decline with age.  In the single degenerate channel, SNe
Ia at 10\,Gyr should be rare, since 1\,\msun\ secondaries have small
envelopes to donate and must rely on only the most massive primaries
\citep{gre05}. A power law DTD (i.e., $\Psi(t)\propto t^{-\beta}$ with
$\beta\sim1$) with a low time-delay cut-off is expected in the double
degenerate scenario \citep[e.g.][]{gre05,for06,mao10}, and has been
explained post-hoc in the single degenerate channel using a mixture of
contributions \citep{hac08}. Furthermore, models with $\beta\sim1$
seem to provide a good match to a variety of recent observational data
\citep{tot08,mao10,gra11}.

We fit both $\beta=1$ and free $\beta$ DTDs to the SNLS+external \snr\
data. The results can be found in Table~\ref{tab:ABchi2} and
Fig.~\ref{fig:rates_dtd_compare}. Our best-fit value is
$\beta=0.98\pm0.05$ ($\chi^2_{\nu}=0.76$), consistent with $1$. This
broad agreement with $1$ holds when considering the other SFH
paramterizations ($\beta=1.15\pm0.08$ for the Cole et al. form).

\citet*[][hereafter PHS]{pri08} present a simple model relating white
dwarf formation rate, which decreases with time following an
instantaneous burst of star formation as $\sim t^{-0.5}$, resulting in
a DTD with $\beta\sim0.5$. By fitting the SN Ia host galaxy data of
\citetalias{sul06b}, \citetalias{pri08} demonstrate that $\Psi(t)\sim
t^{-0.5\pm0.2}$, irrespective of the assumed SFH or the detailed
mixture of stellar populations.  \citetalias{pri08} argue that the
single-degenerate formation scenario alone is not sufficient to
account for all of the observed SNe~Ia \citep[see also][]{gre05}. The
\citetalias{pri08} model makes an explicit prediction for the
evolution of the SN Ia rate with redshift, given an input SFH. We fit
the \citetalias{pri08} model -- essentially $\beta=0.5$ -- to the data
and show the resultant fit in Fig.~\ref{fig:rates_dtd_compare} (also
Table~\ref{tab:rates}). We find a $\chi^2_{\nu}=4.08$, obviously a
substantially poorer fit than the generic power-law fit, or power-law
DTDs with $\beta=1$.

For completeness we also also test an exponential DTD, i.e.,
$\Psi(t)\propto \exp^{-t/\tau}$. When $\tau$ is small, this
approximates a simple star-formation dependence, and when large, it
approximates a constant DTD. The results are in
Table~\ref{tab:ABchi2}; generally, single exponential DTDs provide
poor fits to the data, but do still have acceptable $\chi^2$.

\subsubsection{``Two-component'' models}
\label{sec:aplusb}

Finally, we examine various two-component DTD models. The first is the
popular ``$A+B$'' model,
a simple, two-component model of SN~Ia production that is comprised of
a ``prompt'' component that tracks the instantaneous SFR, and a
``delayed'' (or ``tardy'') component that is proportional to
\mstellar\ \citep{man05,sb05}:
\begin{equation}
  \snr(z) = A\times\mstellar (z) + B\times\mathrm{SFR}(z)
\label{eq:aplusb}
\end{equation}
Here, the $A$ and $B$ coefficients scale the \mstellar\ and SFR
components, respectively.  The prompt component consists of very young
SNe~Ia that explode relatively soon (in the model, immediately) after
the formation of their progenitors, whereas the delayed component
(scaled by $A$) corresponds to longer delay times and an underlying
old stellar population. This model is empirically attractive due to
the ease of comparison with readily observable galaxy quantities, such
as \mstellar\ and SFR.  Note that this A+B model does not exactly
correspond to a DTD -- however, it can be easily converted to a DTD by
using the variation of \mstellar\ with time in a SSP. This leads to a
DTD with some fraction of SNe Ia formed immediately (the $B$
component), followed by a slightly decreasing fraction to large times
(the $A$ component). This decrease is $\sim25$\% from 0.1 to 5\,Gyr,
and $\sim20$\% from 1\,Gyr to 10\,Gyr -- clearly significantly
shallower than a $\beta=1$ power law.

Some confusion exists over the exact definition of \mstellar\ in
eqn.~(\ref{eq:aplusb}), and hence the definition of $A$. Some authors
\citep[e.g.][]{nei06} simply treat \mstellar\ as the integral of the
SFH, equating it to the total formed mass, \mstar. Others make
corrections for stars that have died, particularly in studies which
perform analyses on a galaxy-by-galaxy basis as this quantity is more
straight forward to link to observational data
\citep[e.g.,][]{sul06b}.  This latter definition leads to larger $A$
values, as $\mstellar(t)$ will be less than $\mstar(t)$
\citepalias[see Fig.~7 of][for the size of this difference]{sul06b}.
Here, our $A$ values refer to \mstellar, and we pass the cosmic SFH
through the P\'EGASE.2 routine \citep{leb04}, convolving the chosen
SFH with a single stellar population, and generating a galaxy SED$(t)$
from which mass $\mstellar(t)$ can be estimated. The evolving
\mstellar\ and SFR are used to perform a fit of
equation~\ref{eq:aplusb} to the volumetric rate evolution, with
results listed in Table~\ref{tab:ABchi2}.

%

Fitting the SNLS rates alone gives coefficients of $A=(1.6\pm
0.5)\Aunit$ and $B=(3.4\pm 0.3)\Bunit$ for the \citet{li08} piece-wise
SFH, with $\chi^2_{\nu}=0.74$ for $\nu=7$.  Incorporating the external
SN~Ia rate from LOSS, SDSS and SDF yields values of $A=(1.9\pm
0.1)\Aunit$ and $B=(3.3\pm 0.2)\Bunit$, with $\chi^2_{\nu}=0.60$.
This fit is compared to other DTDs in
Fig.~\ref{fig:rates_dtd_compare}.

Next, we set $A=0$ to investigate the possibility of a pure
star-formation dependence, fitting only the prompt ($B$) component to
the SNLS data.  This results in an upper limit of $B=(4.4\pm
0.3)\Bunit$ ($\chi^2_{\nu}=1.81$), equivalent to the normalized SFH
curve plotted in Fig.~\ref{fig:rates_comp}.  Adding the external data
rates to the SNLS values gives an upper limit of $B=(5.3\pm
0.4)\Bunit$, but again with a very poor fit quality
($\chi^2_{\nu}=6.22$).  While the SNLS results themselves are
marginally consistent with a pure prompt component
(Table~\ref{tab:ABchi2}), adding the additional constraints supplied
by the low-$z$ data yield a very poor $A=0$ fit. The related test of
setting $B=0$ and testing for only the delayed component similarly
gives very poor fit results (Table~\ref{tab:ABchi2}).

As discussed above, these $A$ and $B$ values depend on the adopted
SFH. Even ignoring the systematics in the individual SFR measurements
to which the SFH model is fit, the type of fit used also introduces
considerable uncertainty.  Using the Cole et al. form of the SFH, for
example, changes the best-fit values to $A=1.5\pm0.2\Aunit$ and
$B=4.3\pm0.3\Bunit$, a significant variation with similar quality fit
than for the piece-wise form. Thus care must be taking in comparing
any particular $A+B$ values, or any particular prediction of SN Ia
rate evolution, without ensuring the consistent use of a SFH and
derivation of \mstellar.

Clearly, and as well documented in the literature, the $A+B$ model
must be a significant approximation to the physical reality in SN Ia
progenitor systems. In particular, there must be \textit{some} delay
time for the prompt component, and it is unlikely to act as a delta
function in the DTD. We test this by approximating $\Psi(t)$ as two
discrete bins in time, i.e., a step function with a value $\Psi_1$ at
times $t<t_{\mathrm{split}}$, and $\Psi_2$ at $t\geq
t_{\mathrm{split}}$. We choose $t_{\mathrm{split}}=420$\,Myr
(following, e.g., \citealt{bra10}), and used a sigmoid function to
ensure the DTD was continuous when crossing $t_{\mathrm{tsplit}}$.
This DTD also provides a good fit to the data
(Table~\ref{tab:ABchi2}), with a significant detection of the two
components -- $\Psi_1$ and $\Psi_2$ are $>0$ at 5$\sigma$ in all three
SFHs considered (and typically $\sim10\sigma$). We experimented with
making this function more general, by allowing $t_{\mathrm{split}}$ to
vary. However, these fits were not constraining, although they prefer
$t_{\mathrm{split}}\lesssim2$\,Gyr, and the fit parameters are highly
correlated (i.e., as $t_{\mathrm{tsplit}}$ decreases, $\Psi_1$
increases to ensure a similar fraction of SNe Ia are generated from
the ``prompt'' component).

One direct outcome of this simple two-bin DTD model is that, for our
default cosmic SFH, while $\sim70$\% of SNe Ia originate from the
prompt component integrated over cosmic time, at $z=0$ the prompt
component accounts for only $\sim25$\% of SNe Ia.  These fractions
remain fairly constant out to $t_{\mathrm{split}}\sim2$\,Gyr.

We also explored more bins in $\Psi$ -- e.g., the three bin DTD of
\citet{bra10} and \citet{mao11} -- but, while these were consistent
with our data, they did not provide improved fits over the two-bin
DTD, and again, the parameters themselves were not well constrained.

\subsection{Discussion}
\label{sec:comparison-other-dtd}

\begin{figure}
\plotone{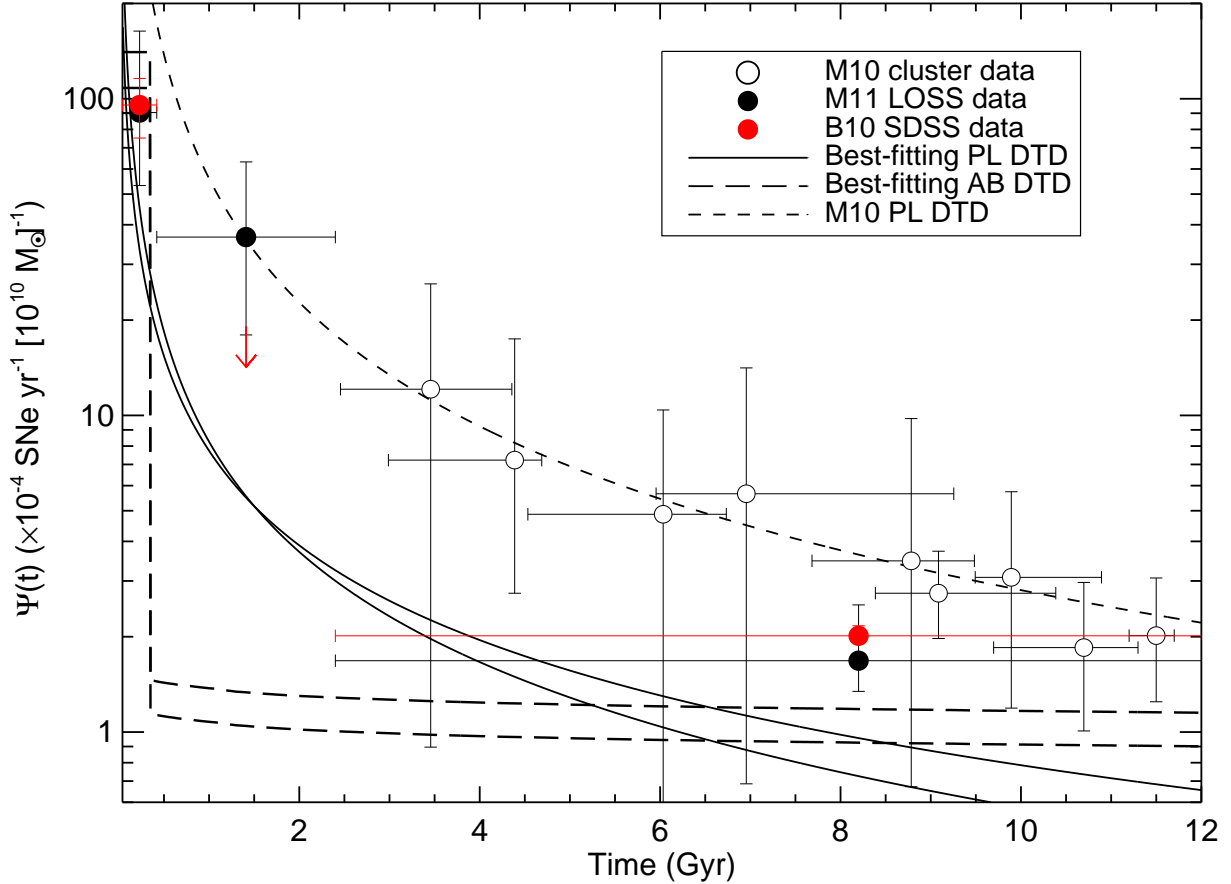}
\caption{The SN Ia delay-time distribution (DTD) inferred from fits to
  our volumetric rate data, compared to other determinations from the
  literature. The solid line shows the best-fitting power-law (PL) DTD
  and the long-dashed line the best-fitting ``$A+B$'' (AB) model from
  Table~\ref{tab:ABchi2}, each drawn for the two SFHs that give the
  most different results. The short-dashed line is the best-fit
  power-law DTD from \citet{mao10}, which has $\beta=1.3$, but with
  their normalization. The $A+B$ model has been adjusted, for plotting
  purposes, so that the instantaneous component is spread over the
  first 400\,Myr. The DTD data points come from \citet[][red
  circles]{bra10}, \citet[][open circles]{mao10}, and \citet[][black
  circles]{mao11}. The horizontal error-bars indicate the bin widths
  on these points.}
\label{fig:dtd}
\end{figure}

We now compare our DTDs inferred from the volumetric \snr\ data with
other independent estimates from the literature, comparing both the
parametric form of our best-fit DTDs, as well as the normalization. In
Fig.~\ref{fig:dtd}, we plot our inferred DTDs, with normalizations
from the best-fits to the volumetric SN Ia rates, and compare to other
empirical determinations of the DTD from the literature from a variety
of methods \citep{bra10,mao10,mao11}. The data points come from the
analysis of SN Ia rates in galaxy clusters \citep{mao10}, the
reconstruction of the DTD from an analysis of the LOSS SN Ia host
galaxy spectra \citep{mao11}, and a similar analysis of the host
galaxies of SDSS SNe Ia \citep{bra10}.

Where necessary we convert these external measurements to a Salpeter
IMF -- from a diet-Salpeter IMF \citep{bel03} for \citet{mao10,mao11}
and a \citet{kro07} IMF for \citet{bra10}. We also adjust the
\citet{mao10} and \citet{mao11} results downwards as their
measurements presumably include SN1991bg-like events which are not
included in our SNLS analysis (the Brandt et al. analysis does not
include these SNe, and their stretch range is well matched to our
analysis -- see their figure 2). Furthermore, we correct the
\citet{bra10} DTD points upwards by 0.26\,dex to account for stellar
masses in Brandt et al. that are 0.26\,dex too high due to a
normalization issue in the VESPA code used to derive them (D. Maoz,
private communication; R. Tojeiro, private communication).

Although the generic shape of the power-law DTD inferred from the
volumetric rate data matches the external DTD data well
(Fig.~\ref{fig:dtd}), it is clear that the best-fit normalization
required to reproduce the volumetric rate data differs to that
required to fit some of the external samples.  The DTDs inferred from
volumetric rate data are generally consistent with the \citet{bra10}
analysis, and the first and third bins of the \citet{mao11} data.

However, the normalization of the best-fit DTD to the \citet{mao10}
cluster data lies significantly above our best-fit DTD.  Integrating
our best-fit power-law DTDs gives
$N_{\mathrm{Ia}}/\mstar\sim4.4\pm0.2-5.2\pm0.2\times10^{-4}\,\mathrm{SNe}\,\msun^{-1}$
($\eta=2.0-2.5$\%) depending on the SFH, in good agreement with
similar analyses \citep{hb10}. However, this is significantly below
the value of $\sim40\times10^{-4}\,\mathrm{SNe}\,\msun^{-1}$ obtained
by integrating the \citet{mao10} ``optimal iron-constraint'' power-law
DTD (for our IMF), or $\sim24\times10^{-4}\,\mathrm{SNe}\,\msun^{-1}$
from the ``minimal iron constraint'' DTD. We can sanity check our
normalizations by predicting the SN Ia rate in the Milky Way, given
our DTD values. Assuming a Milky Way stellar mass of
$\sim5\times10^{10}\msun$ and a SFR of $\sim4\msun\mathrm{yr}^{-1}$,
our $A+B$ DTDs give a predicted Milky Way normal SN Ia rate of
0.22-0.25 events per century. This is in good agreement with
independent estimates of the actual rate \citep[$\sim0.35$ to $0.40$
events per century; e.g.,][]{tam94,rol06,li11a} given the
uncertainties involved. 

So why is the normalization in the \citet{mao10} cluster SN Ia DTD
(which is not arbitrary) different to those derived from volumetric
rate data by such a large factor? Several possibilities exist that may
explain the discrepancy. The first is that the SNLS, and by extension
all other \snr\ studies, are missing a significant number of SNe Ia, a
factor of at least four.  However, it is difficult to understand how
this might occur, given some of the cluster rates used in
\citet{mao10} are drawn from very similar surveys (including SNLS and
SDSS-SN).

A second possibility is that the cosmic SFH models used in our
analysis over-predict the actual SFR -- a lower SFH normalization
would require a higher DTD normalization to match the volumetric rate
data. To test this possibility, we repeat our rates fitting analysis
using the SFH of \citet{wil08}, a SFH derived from a requirement to
match the redshift evolution of \mstellar. Although this agrees with
other SFH estimates at $z<0.7$, it suggests that high-redshift SFRs
could be $\sim0.6$\,dex lower compared to the models used in this
paper.  However, even using this SFH, the integrated power-law DTD
only gives $6.6\pm0.6\times10^{-4}\,\mathrm{SNe}\,\msun^{-1}$
($\eta=3.1\pm0.3$\%), still some distance short of that apparently
required by the clusters analysis.

Other options to adjust the SFH normalization downward, such as
reducing the dust extinction corrections applied to the various
star-formation indicators which make up the SFH compilations, are
probably not viable given the long-established and significant
evidence for obscuration in star-forming galaxies, and the agreement
between the different diagnostics \citep[see discussion
in][]{hb06,li08}.

A third possibility is that the cluster rates used to derive the DTD
of \citet{mao10} have some contamination from ``younger'' SNe Ia, thus
increasing their rates above that appropriate for their age, assuming
a redshift of formation of 3. This, and other similar potential
systematics from the clusters analysis, are discussed in detail in
\citet{mao10}.

Finally, it may well be the case that the assumption of a single DTD
is not adequate, given the various indications that there may be more
than one progenitor channel. This would suggest that there is not one
universal DTD that is independent of redshift, or other variables,
such as metallicity.

\section{Stretch dependence}
\label{sec:stretchdep}

There is an observed correlation between the photometric properties of
SNe~Ia and their host environments \citep{ham95,ham00,sul06b}.
Brighter SNe~Ia with slower light-curves tend to originate in
late-type spiral galaxies, such that the rates of higher-stretch
objects are proportional to star-formation on short ($\sim 0.5$ Gyr)
timescales \citepalias{sul06b}.  Meanwhile, fainter, faster-declining
SNe~Ia are more likely to be associated with older stellar
populations. This split seems to extend to the recovered DTDs --
\citet{bra10} show that the recovered DTD for low and high stretch SNe
Ia are very different, consistent with the above picture of young SNe
Ia being high stretch and old SNe Ia low stretch. A larger fraction of
high-stretch SNe Ia might therefore be expected at high redshift
tracking the increase in the cosmic SFH and hence preponderance of
younger stars.  Indeed, \citet{how07} find a modest increase in the
average light-curve width out to redshifts of $z\sim 1.25$. The rate
evolution for low-stretch SNe~Ia should therefore demonstrate a
correspondingly shallower increase with redshift than that of
higher-stretch objects.

\begin{figure}
\plotone{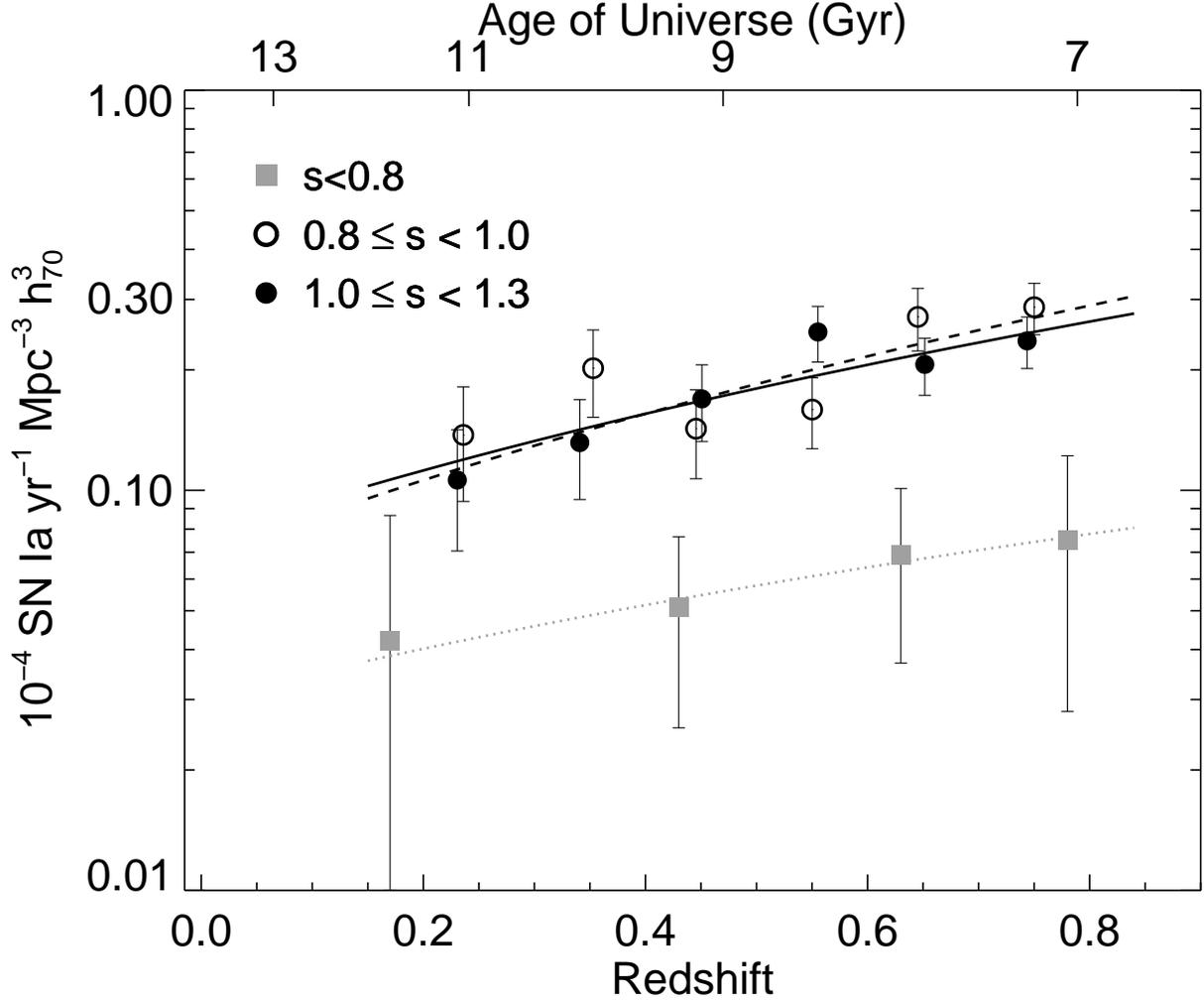}
\caption{SN~Ia rates split at the median SNLS stretch value of
  $s_0=1.0$.  The lower-stretch rates are shown as open circles, and
  high-stretch rates by the solid circles.  We additionally show the
  $s<0.8$ rate measurement from \citet{gon11}. The measured rates have
  simple weighted errors and are uncorrected for systematic errors in
  redshift.  The first two redshift bins ($z=0.15,0.25$) are combined
  here due to the small number of low-$z$ objects the split samples.
  Power-law fits to the rates yield similar slopes over the redshift
  range shown: $\alpha_{s<0.8}=1.63\pm0.28$ (dotted line),
  $\alpha_{0.8\leq s<1}=2.48 \pm 0.97$ (dashed line) and
  $\alpha_{s\geq1}=2.11 \pm 0.60$ (solid line).}
\label{fig:scomp}
\end{figure}

We investigate this trend in Fig.~\ref{fig:scomp}, splitting the SNLS
sample at the median stretch value of $s_0=1.0$. The measured rates
with simple weighted errors are plotted separately for objects with
$0.8 \leq s < 1.0$ (open circles) and those with $1.0\leq s < 1.3$
(filled circles). We also show the $s<0.8$ rate measurement from
\citet{gon11}.  The samples from the analysis in this paper exhibit a
comparable rise in their rates with redshift, with power-law slopes of
$\alpha_{0.8\leq s<1}=2.48 \pm 0.97$ and $\alpha_{s\geq1}=2.11 \pm
0.60$. Extrapolating the fits to each of the samples back to $z=0$
gives the following fractions of SNe Ia at $z=0$ in each group: 17\%
($s<0.8$), 39\% ($0.8 \leq s < 1.0$), and 44\% ($1.0\leq s < 1.3$).
The $s<0.8$ fraction is consistent with the LOSS SN1991bg-like
fraction of 15\% \citep{li11a}, given that true 1991bg-like events
have fitted stretches $s\lesssim0.7$ \citep{gon11}.  The fraction of
very low stretch events with $s<0.8$ SNe Ia shows only a small
increase with increasing redshift \citep[see also][]{gon11}, although
the uncertainties are very large.

The stretch-split rates are only considered out to $z=0.8$, beyond
which redshift the stretch errors become large ($>0.1$), and the lower
efficiencies of the redder, lower-stretch SNe~Ia can bias the observed
stretch evolution by driving up the $s<1$ rates
(Fig.~\ref{fig:objeffs}).  

\begin{figure}
\plotone{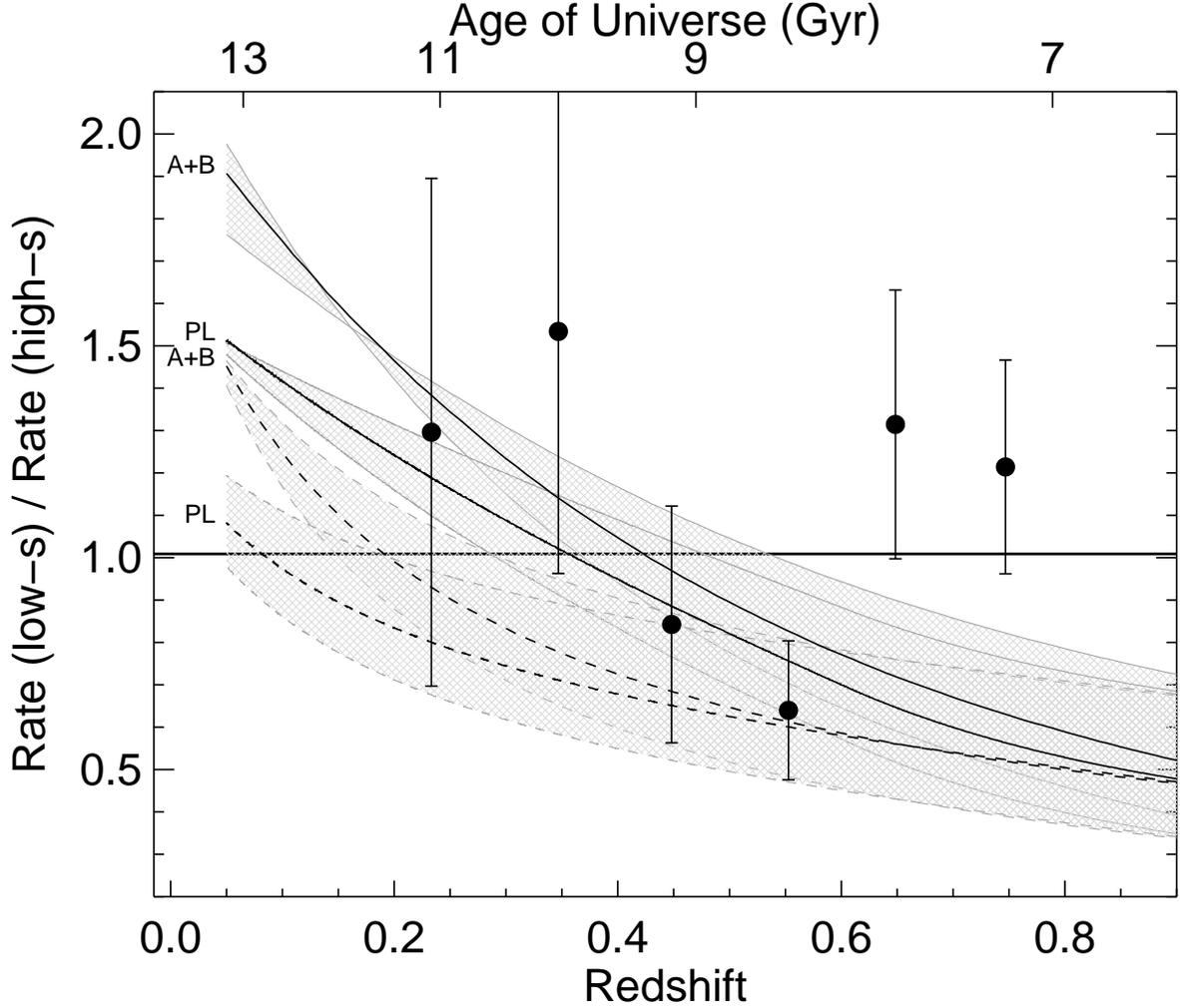}
\caption{The ratio of low-stretch ($0.8<s\leq1$) to high-stretch
  ($1.0<s\leq1.3$) SNe Ia volumetric rates as a function of redshift.
  The horizontal line shows the weighted mean ratio of all SN Ia
  stretches in the sample of $1.01$. Various predicted trends,
  following the analysis of \citet{how07} and using the $A+B$ and
  power-law DTDs, are overlaid.  These trends are not fit to the
  data-points plotted. The solid lines represent the piece-wise SFH,
  and the dashed lines the Cole et al.  form of the SFH, from
  \citet{li08}. The shaded areas show, for each predicted trend, the
  uncertainty expected by shifting the stretch distributions of
  \citet{how07} by 0.025.  The reduced $\chi^2$ of the model fits to
  the data range from $\chi^2_{\nu}=1.8$ to $\chi^2_{\nu}=2.6$ over
  the redshift range shown, compared with $\chi^2_{\nu}=1.35$ for a
  flat line at the weighted mean ratio.}
\label{fig:sratio}
\end{figure}

Fig.~\ref{fig:sratio} shows the ratio of the rates split by stretch
for $s>0.8$.  To compare this observed data with any expected
evolution, we need model stretch distributions for ``old'' and
``young'' SNe, together with a mechanism for predicting the relative
evolution of these two components with redshift. For the former, we
take the two stretch distributions for young and old SNe Ia from
\citet{how07}: the old component is represented by a Gaussian with
$\langle s_{\mathrm{old}} \rangle=0.945$ and
$\sigma_{s_{\mathrm{old}}}=0.077$, while the young component has
$\langle s_{\mathrm{young}} \rangle=1.071$ and
$\sigma_{s_{\mathrm{young}}}=0.063$.  To estimate the relative
redshift evolution of the old and young components, we use the
best-fitting A+B values from Table~\ref{tab:ABchi2} (assigning $A$ to
the old SNe and $B$ to the young SNe), as well as the power-law DTD.
For this latter DTD, we assign SNe born at $t<2$Gyr in the DTD to
the young component, and SNe born at $t\geq2$Gyr to the old component.
Together with a SFH, these models then predict the relative fraction
of low and high stretch SNe as a function of redshift, over-plotted in
Fig.~\ref{fig:sratio}. We also vary the \citet{how07} stretch
distributions by adjusting $\langle s \rangle$ by $\pm0.025$ for the
two components; these are shown as the hashed gray areas in the
figure.

As expected, the predicted ratios show a smooth decline from a large
fraction of low-stretch SNe at $z=0.1$. As the relative contribution
from delayed SNe~Ia to the rates decreases with increasing redshift,
so too should the dominance of lower-stretch objects \citep[see also
Fig.~1 in][]{how07}. The prediction based on the power-law DTD shows a
shallower evolution with redshift, reflecting the extended age of the
young component relative to the simplistic $A+B$ model. 

However, Fig.~\ref{fig:scomp} and Fig.~\ref{fig:sratio} are
surprising. If broad-lightcurve SNe Ia favor a young environment and
narrow-lightcurve events favor an old environment, as has been well
established, then the ratio of narrow to broad SNe ought to be
changing as star formation increases with redshift. But all of the
predictions are a relatively poor match to the SNLS sample in the
higher redshift bins. We vary $s_0$ by $\pm5\%$ to assess the
sensitivity of our results to the stretch split value (default of
1.0), but find no significant improvement in the agreement with the
predicted model as compared with the straight-line fit at the weighted
mean ratio.

The lack of an observed evolution may be due to several factors. The
first is that the age-split between low- and high-s SNe Ia may be more
subtle than previously appreciated. Alternatively, the main effect may
be dominated by $s<0.8$ SNe Ia, which do show a different evolutionary
trend with redshift (Fig.~\ref{fig:scomp}). The limited time baseline
of only $\sim4.5$\,Gyr from $z=0.2-0.8$ may also be a factor, and of
course limitations of the method (e.g., the arbitrary cutoff of for
SNe to be ``young'' or ``old'' in the power-law DTD) could mask any
real effect.

Some other aspects of Fig.~\ref{fig:sratio} are better understood.
The $A+B$ model over-predicts low-s SNe at $z=0$ because it has only
20\% fewer SNe in the DTD at 12\,Gyr compared to 1\,Gyr, in apparent
contrast to the data and to the $t^{-1}$ model, where the rate falls
by an order of magnitude over this baseline. These excess old SNe Ia
from $z=1.5$ show up after a 10 Gyr delay at $z=0$. However, the
difference in the predictions between the $A+B$ and power-law models
is not large, so this unphysical assumption cannot provide the only
solution for the lack of observed evolution in Fig.~\ref{fig:sratio}.

\section{Summary}

In this paper, we have probed the volumetric rate evolution of
``normal'' $0.8<s<1.3$ SNe Ia using a sample of $691$ events from the
Supernova Legacy Survey (SNLS) in the range $0.1<z<1.1$, $286$ of
which have been confirmed spectroscopically.  The SNLS rates increase
with redshift as (1+$z$)$^{\alpha}$ with $\alpha={2.11\pm 0.28}$, and
show no evidence of flattening beyond $z\sim 0.5$.  Due to
spectroscopic incompleteness and the decrease in detection efficiency
for the SNLS sample, a rollover in the slope cannot be ruled out
beyond $z\sim 1$ based on the SNLS data alone.

As a significant component of the SN~Ia rate is linked with young
stellar populations, an increasing fraction of SN~Ia events may suffer
the effects of host extinction at higher redshifts.  In our rate
calculation method, the effect of SN color is factored directly into
the detection efficiency determinations: detection recovery is
evaluated empirically according to the observed SN color regardless of
its cause.  Redder objects at a given redshift have lower detection
efficiencies, and are correspondingly more heavily weighted in the
rates determination.

Combining the SNLS data with that from other SN Ia surveys, we fit
various simple delay-time distributions (DTDs) to the volumetric SN Ia
rate data. DTDs with a single Gaussian are not favored by the data. We
find that simple power-law DTDs ($\Psi(t)\propto t^{-\beta}$) with
$\beta\sim1$ ($\beta=0.98\pm0.05$ to $\beta=1.15\pm0.08$ depending on
the parameterization of the cosmic SFH) can adequately explain all the
SN Ia volumetric rate data, as can two-component models with a
prompt and delayed channel.  These models cannot be separated with
the current volumetric rate data. Integrating these different DTDs
gives the total number of SNe Ia per solar mass formed (excluding
sub-luminous $s<0.8$ events) of
$N_{\mathrm{Ia}}/\mstar\sim4.4-5.7\times10^{-4}\,\mathrm{SNe}\,\msun^{-1}$
(assuming a Salpeter IMF), depending on the star formation history and
DTD model. This is in good agreement with other similar analyses, but
lies significantly below the number expected from DTDs derived from
cluster SN Ia rates.

The use of other techniques, such as fitting the SFH of individual
galaxies \citep{sul06b,bra10,mao11}, or observing a simplified subset
of galaxies \citep{tot08,mao10}, use more information, and in
principle ought to be more reliable. However, each technique has
significant drawbacks, such as contamination \citep{tot08,mao11},
limitations of SED fitting codes \citep{sul06b,bra10,mao11}, and the
assumption that all cluster galaxies formed at $z=3$ in a
delta-function of star-formation \citep{mao11}.  Therefore, our
results are an important complementary constraint. By presenting an
evolution in the SN Ia rate over a large redshift baseline done
self-consistently by a single survey we have for the first time
mitigated the primary drawback of this method -- having to combine
myriad rate determinations from multiple surveys, all done with
different assumptions and biases, sometimes disparate by large factors
\citep{nei06}.

We also find no clear evidence for a difference in the rate evolution
for SNLS samples with $0.8\leq s < 1.0$ and $1.0\leq s< 1.3$ out to
$z=0.8$, although the stretch evolution model from \citet{how07}
cannot be ruled out conclusively.  Stretch evolution plays a more
significant role in the sub-luminous population \citep{gon11}, which
show a much flatter evolution than the $s>0.8$ sample.

Next generation surveys such as Dark Energy Survey (DES), Pan-STARRS,
Palomar Transient Factory (PTF), and SkyMapper, many of which are
already underway, are finding thousands of SNe Ia (in comparison to
the $\sim 700$ in this study).  Statistical rate determinations ought
to improve, but systematic difficulties will remain, as not all SNe
can be spectroscopically confirmed. However, large number statistics
will allow the construction of sub-samples larger than the three (split
by stretch) analyzed here. Comparison of the relative rates of SNe
with different properties and in different environments may ultimately
improve deduced DTDs, and allow for the construction of different DTDs
for subsets of SNe Ia.


\acknowledgments

We are sincerely grateful to the entire Queued Service Observations
team and staff at CFHT for their patience and assistance throughout
the SNLS real-time observing period.  We are particularly indebted to
Pierre Martin, Jean-Charles Cuillandre, Kanoa Withington, and Herb
Woodruff.  Canadian collaboration members acknowledge support from
NSERC and CIAR; French collaboration members from CNRS/IN2P3,
CNRS/INSU and CEA. MS acknowledges support from the Royal Society.

This work is based on observations obtained with MegaPrime/MegaCam, a
joint project of CFHT and CEA/DAPNIA, at the Canada-France-Hawaii
Telescope (CFHT) which is operated by the National Research Council
(NRC) of Canada, the Institut National des Sciences de l'Univers of
the Centre National de la Recherche Scientifique (CNRS) of France, and
the University of Hawaii. This work is based in part on data products
produced at the Canadian Astronomy Data Centre as part of the
Canada-France-Hawaii Telescope Legacy Survey, a collaborative project
of NRC and CNRS.

This work is based in part on observations obtained at the Gemini
Observatory, which is operated by the Association of Universities for
Research in Astronomy, Inc., under a cooperative agreement with the
NSF on behalf of the Gemini partnership: the National Science
Foundation (United States), the Science and Technology Facilities
Council (United Kingdom), the National Research Council (Canada),
CONICYT (Chile), the Australian Research Council (Australia), CNPq
(Brazil) and CONICET (Argentina).  Gemini program IDs: GS-2003B-Q-8,
GN-2003B-Q-9, GS-2004A-Q-11, GN-2004A-Q-19, GS-2004B-Q-31,
GN-2004B-Q-16, GS-2005A-Q-11, GN-2005A-Q-11, GS-2005B-Q-6,
GN-2005B-Q-7, GN-2006A-Q-7, GN-2006B-Q-10, and GN-2007A-Q-8.
Observations made with ESO Telescopes at the Paranal Observatory under
program IDs 171.A-0486 and 176.A-0589.  Some of the data presented
herein were obtained at the W.M. Keck Observatory, which is operated
as a scientific partnership among the California Institute of
Technology, the University of California and the National Aeronautics
and Space Administration. The Observatory was made possible by the
generous financial support of the W.M. Keck Foundation.

\end{document}